\documentclass[manuscript,screen]{acmart}
\PassOptionsToPackage{table}{xcolor}

\usepackage{amsmath}

\usepackage{amssymb}
\usepackage{makecell}
\usepackage{multirow}
\usepackage{multicol}
\usepackage{booktabs}
\usepackage{graphicx}
\usepackage{textcomp}
\usepackage{algorithm}  
\usepackage{algorithmicx}  
\usepackage{algpseudocode}
\usepackage{xspace}
\usepackage{array}
\usepackage{url}
\usepackage[table]{xcolor}
\usepackage{colortbl} 
\usepackage{pifont}
\usepackage{float}
\usepackage{bm}

\AtBeginDocument{%
  \providecommand\BibTeX{{%
    \normalfont B\kern-0.5em{\scshape i\kern-0.25em b}\kern-0.8em\TeX}}}

\setcopyright{acmlicensed}
\copyrightyear{2024}
\acmYear{2024}
\acmDOI{XXXXXXX.XXXXXXX}

\begin{document}

\title{Automatically Learning a Precise Measurement for Fault Diagnosis Capability of Test Cases}

\author{Yifan Zhao}
\email{zhaoyifan@stu.pku.edu.cn}
\affiliation{%
  \institution{Key Laboratory of High Confidence Software Technologies (Peking University), Ministry of Education; School of Computer Science, Peking University}
  \city{Beijing}
  \country{China}
}

\author{Zeyu Sun}
\email{zeyu.zys@gmail.com}
\affiliation{%
 \institution{National Key Laboratory of Space
Integrated Information System, Institute of Software, Chinese Academy of Sciences}
  \city{Beijing}
  \country{China}
}

\author{Guoqing Wang}
\email{guoqingwang@stu.pku.edu.cn}
\affiliation{%
  \institution{Key Laboratory of High Confidence Software Technologies (Peking University), Ministry of Education; School of Computer Science, Peking University}
  \city{Beijing}
  \country{China}
}

\author{Qingyuan Liang}
\email{liangqy@stu.pku.edu.cn}
\affiliation{%
  \institution{Key Laboratory of High Confidence Software Technologies (Peking University), Ministry of Education; School of Computer Science, Peking University}
  \city{Beijing}
  \country{China}
}

\author{Yakun Zhang}
\email{zhangyakun@stu.pku.edu.cn}
\affiliation{%
  \institution{Key Laboratory of High Confidence Software Technologies (Peking University), Ministry of Education; School of Computer Science, Peking University}
  \city{Beijing}
  \country{China}
}

\author{Yiling Lou}
\email{yilinglou@fudan.edu.cn}
\affiliation{%
  \institution{Fudan University}
  \city{Shanghai}
  \country{China}
}

\author{Dan Hao}
\email{haodan@pku.edu.cn}
\authornote{Dan Hao is the corresponding author.}
\affiliation{%
  \institution{Key Laboratory of High Confidence Software Technologies (Peking University), Ministry of Education; School of Computer Science, Peking University}
  \city{Beijing}
  \country{China}
}

\author{Lu Zhang}
\email{zhanglucs@pku.edu.cn}
\affiliation{%
  \institution{Key Laboratory of High Confidence Software Technologies (Peking University), Ministry of Education; School of Computer Science, Peking University}
  \city{Beijing}
  \country{China}
}

\renewcommand{\shortauthors}{Zhao et al.}
\newcommand{\techname}{{RLFDC}\xspace}

\begin{abstract}
Prevalent Fault Localization (FL) techniques rely on tests to localize buggy program elements. Tests could be treated as fuel to further boost FL by providing more debugging information. Therefore, it is highly valuable to measure the Fault Diagnosis Capability (FDC) of a test for diagnosing faults, so as to select or generate tests to better help FL (i.e., FL-oriented test selection or FL-oriented test generation). To this end, researchers have proposed many FDC metrics, which serve as the selection criterion in FL-oriented test selection or the fitness function in FL-oriented test generation. Existing FDC metrics can be classified into result-agnostic and result-aware metrics depending on whether they take test results (i.e., passing or failing) as input. Although result-aware metrics perform better in test selection, they have restricted applications due to the input of test results, e.g., they cannot be applied to guide test generation. Moreover, all the existing FDC metrics are designed based on some predefined heuristics and have achieved limited FL performance due to their inaccuracy. To address these issues, in this paper, we reconsider result-agnostic metrics (i.e., metrics that do not take test results as input), and propose a novel result-agnostic metric \techname{} which predicts FDC values of tests through reinforcement learning. In particular, we treat FL results as reward signals, and train an FDC prediction model with the direct FL feedback to automatically learn a more accurate measurement rather than design one based on predefined heuristics. Finally, we evaluate the proposed \techname{} on Defects4J by applying the studied metrics to test selection and generation. According to the experimental results, the proposed \techname{} outperforms all the result-agnostic metrics in both test selection and generation, e.g., when applied to selecting human-written tests, \techname{} achieves 28.2\% and 21.6\% higher acc@1 and mAP values compared to the state-of-the-art result-agnostic metric TfD. Besides, \techname{} even achieves competitive performance compared to the state-of-the-art result-aware metric FDG in test selection.
\end{abstract}



\received{14 March 2024}
\received[revised]{26 November 2024}
\received[accepted]{17 December 2024}

\maketitle

\section{Introduction}
\label{sec:intro}
Software debugging is a critical but painstaking process that costs tremendous resources every year~\cite{vessey1985expertise,wong2016survey}. To alleviate this problem, Fault Localization (FL) techniques aim to automatically localize the buggy elements~\cite{wong2016survey} and thus reduce the cost. Prevalent FL techniques~\cite{naish2011model,abreu2009practical,jones2005empirical,wong2013dstar,lou2021boosting,li2021fault,li2019deepfl,zhang2019cnn,papadakis2015metallaxis,hong2015mutation,moon2014ask,papadakis2012using} rely on the coverage information and execution results of tests as input to predict a suspicious value for each code element, indicating their possibility of harboring faults. Consequently, the effectiveness of FL techniques heavily relies on the quality of the tests, as testing information provides critical clues for fault diagnosis. In other words, the fault diagnosis capability (FDC) of a test suite might potentially become the bottleneck of the FL effectiveness~\cite{ddu, entbug}. Unfortunately, in practice, tests with high FDC are not always available, especially considering the heavy expense for manually writing tests~\cite{dinella2022toga}. 

\begin{figure}
    \centering
    \includegraphics[width=0.7\linewidth]{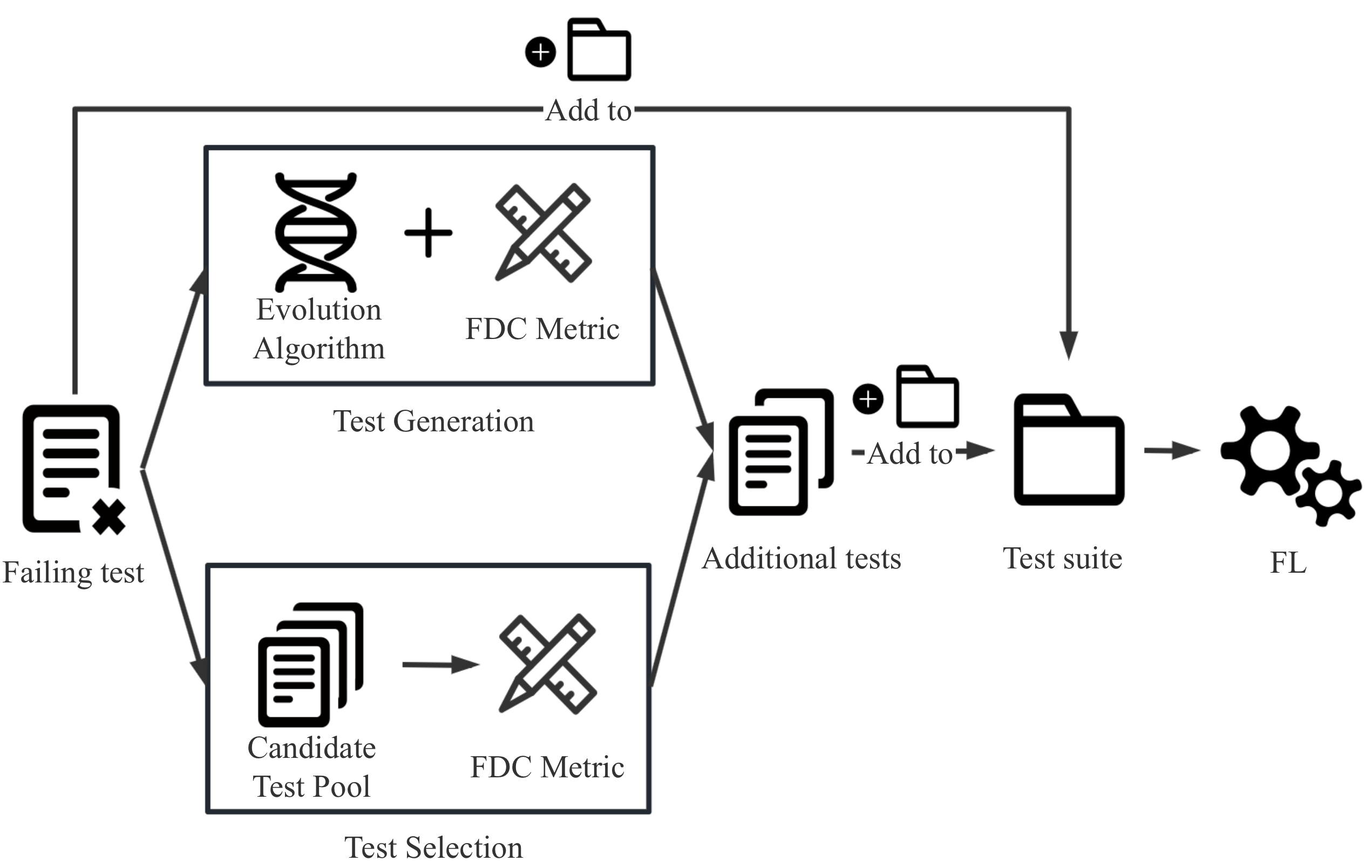}
    \caption{The overall workflow of FDC metric}
    \vspace{-0.2cm}
    \label{fig:intro_overview}
\end{figure}

To mitigate this issue, researchers propose FDC metrics to measure the diagnostic capability of a test or a test suite, and apply these FDC metrics to test generation~\cite{ddu,tfd,entbug} or selection~\cite{hao2010test,an2022fdg} for improving FL. 
Fig.~\ref{fig:intro_overview} presents the workflow of these applications, where FDC metrics play a central role. Given a fault-triggering test (which composes an initial test suite), both of these two tasks aim to augment the initial test suite by generating or selecting more additional tests to more accurately pinpoint the fault. In particular, test generation for FL aims to automatically generate more high-quality tests. It utilizes FDC metric as a fitness function to guide the evolutionary algorithm of test generation tools and generate more tests with high FDC. These tests are then used to construct an augmented test suite to boost the performance of FL techniques. Test selection for FL aims to select valuable tests with high FDC from a given test pool. The selected tests are then used to construct an augmented test suite to boost FL. Selection is performed here because the existing test generation tools cannot generate accurate test oracles and thus the oracles are usually manually labelled~\cite{entbug}. To alleviate human efforts on oracle labeling, FDC metrics are applied to select a subset of valuable tests from the whole set of tests generated by test generation tools. The selected tests are then labeled with oracles and provides additional debugging information for FL techniques. 
In summary, FDC metric serves as a fitness function in test generation and a selection criterion in test selection. Generally, test-based FL techniques, especially spectrum-based FL~\cite{ochiai,abreu2009practical},
predict suspicious values of code elements based on the diagnosis capability of tests, and FDC metric is proposed to quantify this capability. 

Existing FDC metrics 
can be classified into result-agnostic~\cite{ddu,tfd,entbug,gonzalez2011prioritizing} and result-aware metrics~\cite{an2022fdg,bandyopadhyay2011proximity,hao2010test,yoo2013fault}. Result-agnostic metrics calculate the FDC value based on test coverage, while result-aware metrics calculate the FDC value based on test coverage and previous test execution results. The extra input of execution results makes result-aware metrics perform better but limits their application scenarios. For example, they cannot be applied to test generation for FL, because during the evolution process in test generation, accurate oracles cannot be automatically generated~\cite{ddu,entbug}. 
Moreover, all existing metrics 
are based on hand-crafted formulas and therefore use fixed heuristics on all subjects. However, the prior knowledge may be inaccurate and lacks general applicability for different subjects, making the proposed metrics inherently ineffective. 
For example, our experimental results on Defects4J show that the result-agnostic metric TfD~\cite{tfd} performs well on \textit{Chart} and \textit{Lang}, but performs badly on \textit{Time} and \textit{Math}. 


To solve this problem, we propose a Reinforcement Learning (RL) based approach to build a more accurate result-agnostic FDC metric, \techname{}.
Note that, differing slightly from previous work~\cite{an2022fdg}\footnote{Prior work~\cite{an2022fdg} classify metrics as ``result-aware'' or ``result-agnostic'' based on whether they require test execution results.}, we classify metrics as ``result-aware'' or ``result-agnostic'' based on whether they require execution results from tests other than the initial failing test. Our focus is on localizing the fault for a given failing test, thus the information from this initial test is known and can be leveraged. Importantly, utilizing this initial information does not impede the application of RLFDC to test generation; therefore, we categorize it as a result-agnostic metric.
Different from previous metrics built with ingenious formulas, an RL model is built through repetitive training to automatically learn a measuring strategy, and the finalized RL model serves as our FDC metric \techname{}. Concretely, during the training stage, our approach uses the coverage information of tests to calculate state and action in RL, and uses the actual FL results to calculate the reward to motivate the RL model. Following Fig.~\ref{fig:intro_overview}, we investigate the performance of \techname{} in test selection and generation for FL, which can be regarded as the evaluating stage of the RL model.
In test generation for FL, we use an FDC metric as the fitness function of an evolutionary algorithm and thus integrate it with a popular test generation tool EVOSUITE~\cite{fraser2011evosuite}. In test selection for FL, we use an FDC metric as the selection criterion and apply it to a test pool.


We evaluate \techname{} on the widely-used benchmark Defects4J~\cite{just2014defects4j}, by comparing it against three representative result-agnostic metrics (i.e., EntBug~\cite{entbug}, DDU~\cite{ddu}, and TfD~\cite{tfd}) and the state-of-the-art result-aware metric FDG~\cite{an2022fdg}. We evaluate the FDC metrics in three FL-oriented scenarios: test selection in human-written tests, test selection in automatically generated tests, and automated test generation. In each scenario, FDC metrics are used to guide test selection or generation, and their performance is evaluated via feeding the selected or generated tests to an FL technique. According to the results, in the test selection scenario, \techname{} outperforms all the existing result-agnostic metrics, while achieving competitive performance with the state-of-the-art result-aware metric FDG. With human-written tests, \techname{} achieves 28.2\%, 36.2\% and 21.6\% higher acc@1, acc@10, and mAP (Mean Average Precision) values compared to the existing best result-agnostic metric TfD~\cite{an2022fdg}, respectively. 
With tests automatically generated by EVOSUITE, RLFDC achieves 3.0\%, 5.6\% and 4.4\% higher acc@1, acc@10, and mAP values compared to TfD, respectively. With tests automatically generated by Randoop~\cite{pacheco2007randoop}, RLFDC achieves 12.0\%, 8.1\% and 3.4\% higher acc@1, acc@10, and mAP values compared to TfD, respectively.
In the test generation scenario where result-aware metrics cannot be applied, EVOSUITE under the guidance of \techname{} successfully generates tests with high FDC values and helps improve FL performance. In particular, \techname{} achieves 4.9\% and 3.1\% higher acc@1 and mAP values compared to the state-of-the-art FL-oriented test generation technique DDU, respectively. We also evaluate \techname{} in the cross-project scenario and the results show that \techname{} steadily outperforms state-of-the-art result-agnostic metrics, indicating its general effectiveness. Finally, we perform an ablation study and the results show that each component of \techname{} positively contributes to its effectiveness.


The contribution of this paper can be summarised as follows:
\begin{itemize}
    \item A new result-agnostic metric \techname{} which is constructed by our RL-based approach. To the best of our knowledge, we are the first to apply RL to this task.
    \item An extensive evaluation of \techname{}, including three scenarios: test selection (based on both human-written and automatically generated tests) and test generation.  
    \item A prototype implementation of a RLFDC-guided test generation tool on top of EVOSUITE~\cite{fraser2011evosuite}. To our best knowledge, we are the first to integrate EVOSUITE with a machine learning model to guide test generation. 
\end{itemize}

\vspace{-0.2cm}
\section{Motivating Example}
\label{sec:example}

\begin{table}[htp]
\centering
\caption{Coverage information for tests selected by RLFDC and TfD}
\vspace{-0.2cm}
\resizebox{1.0\linewidth}{!}{
\begin{tabular}{|cc|c|cccccccccc|cccccccccc|}
\hline
\multicolumn{1}{|l}{}                                             & \multicolumn{1}{l|}{}                                & \begin{tabular}[c]{@{}c@{}}Init.\\ Test\end{tabular} & \multicolumn{10}{c|}{Tests selected by RLFDC}                                                                                                                                                                                                 & \multicolumn{10}{c|}{Tests selected by TfD}                                                                                                                                                                                                            \\ \hline
\multicolumn{1}{|c|}{Method}                                      & \begin{tabular}[c]{@{}c@{}}State\\ ment\end{tabular} & t0                                                   & \multicolumn{1}{c|}{t1} & \multicolumn{1}{c|}{t2} & \multicolumn{1}{c|}{t3} & \multicolumn{1}{c|}{t4} & \multicolumn{1}{c|}{t5} & \multicolumn{1}{c|}{t6} & \multicolumn{1}{c|}{t7} & \multicolumn{1}{c|}{t8} & \multicolumn{1}{c|}{t9} & t10 & \multicolumn{1}{c|}{t11} & \multicolumn{1}{c|}{t12} & \multicolumn{1}{c|}{t13} & \multicolumn{1}{c|}{t14} & \multicolumn{1}{c|}{t15} & \multicolumn{1}{c|}{t16} & \multicolumn{1}{c|}{t17} & \multicolumn{1}{c|}{t18} & \multicolumn{1}{c|}{t19} & t20 \\ \hline
\multicolumn{1}{|c|}{m1}                                          & s1                                                   & 1                                                    & \multicolumn{1}{c|}{1}  & \multicolumn{1}{c|}{1}  & \multicolumn{1}{c|}{1}  & \multicolumn{1}{c|}{1}  & \multicolumn{1}{c|}{1}  & \multicolumn{1}{c|}{1}  & \multicolumn{1}{c|}{1}  & \multicolumn{1}{c|}{1}  & \multicolumn{1}{c|}{1}  & 0   & \multicolumn{1}{c|}{1}   & \multicolumn{1}{c|}{1}   & \multicolumn{1}{c|}{1}   & \multicolumn{1}{c|}{1}   & \multicolumn{1}{c|}{0}   & \multicolumn{1}{c|}{0}   & \multicolumn{1}{c|}{1}   & \multicolumn{1}{c|}{0}   & \multicolumn{1}{c|}{0}   & 1   \\ \hline
\multicolumn{1}{|c|}{m2}                                          & s2                                                   & 1                                                    & \multicolumn{1}{c|}{1}  & \multicolumn{1}{c|}{0}  & \multicolumn{1}{c|}{0}  & \multicolumn{1}{c|}{1}  & \multicolumn{1}{c|}{0}  & \multicolumn{1}{c|}{0}  & \multicolumn{1}{c|}{0}  & \multicolumn{1}{c|}{0}  & \multicolumn{1}{c|}{0}  & 0   & \multicolumn{1}{c|}{0}   & \multicolumn{1}{c|}{0}   & \multicolumn{1}{c|}{0}   & \multicolumn{1}{c|}{0}   & \multicolumn{1}{c|}{0}   & \multicolumn{1}{c|}{0}   & \multicolumn{1}{c|}{0}   & \multicolumn{1}{c|}{0}   & \multicolumn{1}{c|}{0}   & 0   \\ \hline
\multicolumn{1}{|c|}{}                                            & s3                                                   & 1                                                    & \multicolumn{1}{c|}{1}  & \multicolumn{1}{c|}{0}  & \multicolumn{1}{c|}{0}  & \multicolumn{1}{c|}{1}  & \multicolumn{1}{c|}{0}  & \multicolumn{1}{c|}{0}  & \multicolumn{1}{c|}{1}  & \multicolumn{1}{c|}{0}  & \multicolumn{1}{c|}{0}  & 0   & \multicolumn{1}{c|}{0}   & \multicolumn{1}{c|}{0}   & \multicolumn{1}{c|}{0}   & \multicolumn{1}{c|}{0}   & \multicolumn{1}{c|}{0}   & \multicolumn{1}{c|}{0}   & \multicolumn{1}{c|}{0}   & \multicolumn{1}{c|}{0}   & \multicolumn{1}{c|}{0}   & 0   \\ \cline{2-23} 
\multicolumn{1}{|c|}{}                                            & s4                                                   & 1                                                    & \multicolumn{1}{c|}{1}  & \multicolumn{1}{c|}{0}  & \multicolumn{1}{c|}{0}  & \multicolumn{1}{c|}{1}  & \multicolumn{1}{c|}{0}  & \multicolumn{1}{c|}{0}  & \multicolumn{1}{c|}{1}  & \multicolumn{1}{c|}{0}  & \multicolumn{1}{c|}{0}  & 0   & \multicolumn{1}{c|}{0}   & \multicolumn{1}{c|}{0}   & \multicolumn{1}{c|}{0}   & \multicolumn{1}{c|}{0}   & \multicolumn{1}{c|}{0}   & \multicolumn{1}{c|}{0}   & \multicolumn{1}{c|}{0}   & \multicolumn{1}{c|}{0}   & \multicolumn{1}{c|}{0}   & 0   \\ \cline{2-23} 
\multicolumn{1}{|c|}{}                                            & s5                                                   & 1                                                    & \multicolumn{1}{c|}{1}  & \multicolumn{1}{c|}{0}  & \multicolumn{1}{c|}{0}  & \multicolumn{1}{c|}{1}  & \multicolumn{1}{c|}{0}  & \multicolumn{1}{c|}{0}  & \multicolumn{1}{c|}{1}  & \multicolumn{1}{c|}{0}  & \multicolumn{1}{c|}{0}  & 0   & \multicolumn{1}{c|}{0}   & \multicolumn{1}{c|}{0}   & \multicolumn{1}{c|}{0}   & \multicolumn{1}{c|}{0}   & \multicolumn{1}{c|}{0}   & \multicolumn{1}{c|}{0}   & \multicolumn{1}{c|}{0}   & \multicolumn{1}{c|}{0}   & \multicolumn{1}{c|}{0}   & 0   \\ \cline{2-23} 
\multicolumn{1}{|c|}{}                                            & s6                                                   & 1                                                    & \multicolumn{1}{c|}{1}  & \multicolumn{1}{c|}{0}  & \multicolumn{1}{c|}{0}  & \multicolumn{1}{c|}{1}  & \multicolumn{1}{c|}{0}  & \multicolumn{1}{c|}{0}  & \multicolumn{1}{c|}{1}  & \multicolumn{1}{c|}{0}  & \multicolumn{1}{c|}{0}  & 0   & \multicolumn{1}{c|}{0}   & \multicolumn{1}{c|}{0}   & \multicolumn{1}{c|}{0}   & \multicolumn{1}{c|}{0}   & \multicolumn{1}{c|}{0}   & \multicolumn{1}{c|}{0}   & \multicolumn{1}{c|}{0}   & \multicolumn{1}{c|}{0}   & \multicolumn{1}{c|}{0}   & 0   \\ \cline{2-23} 
\multicolumn{1}{|c|}{}                                            & s7                                                   & 1                                                    & \multicolumn{1}{c|}{1}  & \multicolumn{1}{c|}{0}  & \multicolumn{1}{c|}{0}  & \multicolumn{1}{c|}{1}  & \multicolumn{1}{c|}{0}  & \multicolumn{1}{c|}{0}  & \multicolumn{1}{c|}{1}  & \multicolumn{1}{c|}{0}  & \multicolumn{1}{c|}{0}  & 0   & \multicolumn{1}{c|}{0}   & \multicolumn{1}{c|}{0}   & \multicolumn{1}{c|}{0}   & \multicolumn{1}{c|}{0}   & \multicolumn{1}{c|}{0}   & \multicolumn{1}{c|}{0}   & \multicolumn{1}{c|}{0}   & \multicolumn{1}{c|}{0}   & \multicolumn{1}{c|}{0}   & 0   \\ \cline{2-23} 
\multicolumn{1}{|c|}{\multirow{-6}{*}{m3}}                        & s8                                                   & 1                                                    & \multicolumn{1}{c|}{1}  & \multicolumn{1}{c|}{0}  & \multicolumn{1}{c|}{0}  & \multicolumn{1}{c|}{1}  & \multicolumn{1}{c|}{0}  & \multicolumn{1}{c|}{0}  & \multicolumn{1}{c|}{1}  & \multicolumn{1}{c|}{0}  & \multicolumn{1}{c|}{0}  & 0   & \multicolumn{1}{c|}{0}   & \multicolumn{1}{c|}{0}   & \multicolumn{1}{c|}{0}   & \multicolumn{1}{c|}{0}   & \multicolumn{1}{c|}{0}   & \multicolumn{1}{c|}{0}   & \multicolumn{1}{c|}{0}   & \multicolumn{1}{c|}{0}   & \multicolumn{1}{c|}{0}   & 0   \\ \hline
\multicolumn{1}{|c|}{{\color[HTML]{FE0000} }}                     & s9                                                   & 1                                                    & \multicolumn{1}{c|}{1}  & \multicolumn{1}{c|}{0}  & \multicolumn{1}{c|}{0}  & \multicolumn{1}{c|}{0}  & \multicolumn{1}{c|}{0}  & \multicolumn{1}{c|}{0}  & \multicolumn{1}{c|}{0}  & \multicolumn{1}{c|}{0}  & \multicolumn{1}{c|}{0}  & 0   & \multicolumn{1}{c|}{0}   & \multicolumn{1}{c|}{0}   & \multicolumn{1}{c|}{0}   & \multicolumn{1}{c|}{0}   & \multicolumn{1}{c|}{0}   & \multicolumn{1}{c|}{0}   & \multicolumn{1}{c|}{0}   & \multicolumn{1}{c|}{0}   & \multicolumn{1}{c|}{0}   & 0   \\ \cline{2-23} 
\multicolumn{1}{|c|}{{\color[HTML]{FE0000} }}                     & s10                                                  & 1                                                    & \multicolumn{1}{c|}{1}  & \multicolumn{1}{c|}{1}  & \multicolumn{1}{c|}{1}  & \multicolumn{1}{c|}{0}  & \multicolumn{1}{c|}{0}  & \multicolumn{1}{c|}{0}  & \multicolumn{1}{c|}{0}  & \multicolumn{1}{c|}{0}  & \multicolumn{1}{c|}{0}  & 0   & \multicolumn{1}{c|}{0}   & \multicolumn{1}{c|}{0}   & \multicolumn{1}{c|}{0}   & \multicolumn{1}{c|}{0}   & \multicolumn{1}{c|}{0}   & \multicolumn{1}{c|}{0}   & \multicolumn{1}{c|}{0}   & \multicolumn{1}{c|}{0}   & \multicolumn{1}{c|}{0}   & 0   \\ \cline{2-23} 
\multicolumn{1}{|c|}{{\color[HTML]{FE0000} }}                     & s11                                                  & 1                                                    & \multicolumn{1}{c|}{1}  & \multicolumn{1}{c|}{1}  & \multicolumn{1}{c|}{0}  & \multicolumn{1}{c|}{0}  & \multicolumn{1}{c|}{0}  & \multicolumn{1}{c|}{0}  & \multicolumn{1}{c|}{0}  & \multicolumn{1}{c|}{0}  & \multicolumn{1}{c|}{0}  & 0   & \multicolumn{1}{c|}{0}   & \multicolumn{1}{c|}{0}   & \multicolumn{1}{c|}{0}   & \multicolumn{1}{c|}{0}   & \multicolumn{1}{c|}{0}   & \multicolumn{1}{c|}{0}   & \multicolumn{1}{c|}{0}   & \multicolumn{1}{c|}{0}   & \multicolumn{1}{c|}{0}   & 0   \\ \cline{2-23} 
\multicolumn{1}{|c|}{{\color[HTML]{FE0000} }}                     & {\color[HTML]{FE0000} s12}                           & 1                                                    & \multicolumn{1}{c|}{1}  & \multicolumn{1}{c|}{1}  & \multicolumn{1}{c|}{0}  & \multicolumn{1}{c|}{0}  & \multicolumn{1}{c|}{0}  & \multicolumn{1}{c|}{0}  & \multicolumn{1}{c|}{0}  & \multicolumn{1}{c|}{0}  & \multicolumn{1}{c|}{0}  & 0   & \multicolumn{1}{c|}{0}   & \multicolumn{1}{c|}{0}   & \multicolumn{1}{c|}{0}   & \multicolumn{1}{c|}{0}   & \multicolumn{1}{c|}{0}   & \multicolumn{1}{c|}{0}   & \multicolumn{1}{c|}{0}   & \multicolumn{1}{c|}{0}   & \multicolumn{1}{c|}{0}   & 0   \\ \cline{2-23} 
\multicolumn{1}{|c|}{{\color[HTML]{FE0000} }}                     & {\color[HTML]{FE0000} s13}                           & 1                                                    & \multicolumn{1}{c|}{1}  & \multicolumn{1}{c|}{1}  & \multicolumn{1}{c|}{0}  & \multicolumn{1}{c|}{0}  & \multicolumn{1}{c|}{0}  & \multicolumn{1}{c|}{0}  & \multicolumn{1}{c|}{0}  & \multicolumn{1}{c|}{0}  & \multicolumn{1}{c|}{0}  & 0   & \multicolumn{1}{c|}{0}   & \multicolumn{1}{c|}{0}   & \multicolumn{1}{c|}{0}   & \multicolumn{1}{c|}{0}   & \multicolumn{1}{c|}{0}   & \multicolumn{1}{c|}{0}   & \multicolumn{1}{c|}{0}   & \multicolumn{1}{c|}{0}   & \multicolumn{1}{c|}{0}   & 0   \\ \cline{2-23} 
\multicolumn{1}{|c|}{{\color[HTML]{FE0000} }}                     & {\color[HTML]{FE0000} s14}                           & 1                                                    & \multicolumn{1}{c|}{1}  & \multicolumn{1}{c|}{0}  & \multicolumn{1}{c|}{0}  & \multicolumn{1}{c|}{0}  & \multicolumn{1}{c|}{0}  & \multicolumn{1}{c|}{0}  & \multicolumn{1}{c|}{0}  & \multicolumn{1}{c|}{0}  & \multicolumn{1}{c|}{0}  & 0   & \multicolumn{1}{c|}{0}   & \multicolumn{1}{c|}{0}   & \multicolumn{1}{c|}{0}   & \multicolumn{1}{c|}{0}   & \multicolumn{1}{c|}{0}   & \multicolumn{1}{c|}{0}   & \multicolumn{1}{c|}{0}   & \multicolumn{1}{c|}{0}   & \multicolumn{1}{c|}{0}   & 0   \\ \cline{2-23} 
\multicolumn{1}{|c|}{{\color[HTML]{FE0000} }}                     & s15                                                  & 1                                                    & \multicolumn{1}{c|}{1}  & \multicolumn{1}{c|}{1}  & \multicolumn{1}{c|}{1}  & \multicolumn{1}{c|}{0}  & \multicolumn{1}{c|}{0}  & \multicolumn{1}{c|}{0}  & \multicolumn{1}{c|}{0}  & \multicolumn{1}{c|}{0}  & \multicolumn{1}{c|}{0}  & 0   & \multicolumn{1}{c|}{0}   & \multicolumn{1}{c|}{0}   & \multicolumn{1}{c|}{0}   & \multicolumn{1}{c|}{0}   & \multicolumn{1}{c|}{0}   & \multicolumn{1}{c|}{0}   & \multicolumn{1}{c|}{0}   & \multicolumn{1}{c|}{0}   & \multicolumn{1}{c|}{0}   & 0   \\ \cline{2-23} 
\multicolumn{1}{|c|}{{\color[HTML]{FE0000} }}                     & s16                                                  & 1                                                    & \multicolumn{1}{c|}{1}  & \multicolumn{1}{c|}{0}  & \multicolumn{1}{c|}{1}  & \multicolumn{1}{c|}{0}  & \multicolumn{1}{c|}{0}  & \multicolumn{1}{c|}{0}  & \multicolumn{1}{c|}{0}  & \multicolumn{1}{c|}{0}  & \multicolumn{1}{c|}{0}  & 0   & \multicolumn{1}{c|}{0}   & \multicolumn{1}{c|}{0}   & \multicolumn{1}{c|}{0}   & \multicolumn{1}{c|}{0}   & \multicolumn{1}{c|}{0}   & \multicolumn{1}{c|}{0}   & \multicolumn{1}{c|}{0}   & \multicolumn{1}{c|}{0}   & \multicolumn{1}{c|}{0}   & 0   \\ \cline{2-23} 
\multicolumn{1}{|c|}{\multirow{-9}{*}{{\color[HTML]{FE0000} m4}}} & s17                                                  & 1                                                    & \multicolumn{1}{c|}{1}  & \multicolumn{1}{c|}{1}  & \multicolumn{1}{c|}{1}  & \multicolumn{1}{c|}{0}  & \multicolumn{1}{c|}{0}  & \multicolumn{1}{c|}{0}  & \multicolumn{1}{c|}{0}  & \multicolumn{1}{c|}{0}  & \multicolumn{1}{c|}{0}  & 0   & \multicolumn{1}{c|}{0}   & \multicolumn{1}{c|}{0}   & \multicolumn{1}{c|}{0}   & \multicolumn{1}{c|}{0}   & \multicolumn{1}{c|}{0}   & \multicolumn{1}{c|}{0}   & \multicolumn{1}{c|}{0}   & \multicolumn{1}{c|}{0}   & \multicolumn{1}{c|}{0}   & 0   \\ \hline
\multicolumn{1}{|c|}{m5}                                          & s18                                                  & 1                                                    & \multicolumn{1}{c|}{1}  & \multicolumn{1}{c|}{1}  & \multicolumn{1}{c|}{1}  & \multicolumn{1}{c|}{0}  & \multicolumn{1}{c|}{1}  & \multicolumn{1}{c|}{1}  & \multicolumn{1}{c|}{0}  & \multicolumn{1}{c|}{0}  & \multicolumn{1}{c|}{0}  & 0   & \multicolumn{1}{c|}{0}   & \multicolumn{1}{c|}{1}   & \multicolumn{1}{c|}{0}   & \multicolumn{1}{c|}{0}   & \multicolumn{1}{c|}{0}   & \multicolumn{1}{c|}{0}   & \multicolumn{1}{c|}{0}   & \multicolumn{1}{c|}{0}   & \multicolumn{1}{c|}{0}   & 1   \\ \hline
\multicolumn{1}{|c|}{}                                            & s19                                                  & 1                                                    & \multicolumn{1}{c|}{1}  & \multicolumn{1}{c|}{1}  & \multicolumn{1}{c|}{1}  & \multicolumn{1}{c|}{0}  & \multicolumn{1}{c|}{1}  & \multicolumn{1}{c|}{1}  & \multicolumn{1}{c|}{0}  & \multicolumn{1}{c|}{1}  & \multicolumn{1}{c|}{1}  & 0   & \multicolumn{1}{c|}{0}   & \multicolumn{1}{c|}{1}   & \multicolumn{1}{c|}{1}   & \multicolumn{1}{c|}{1}   & \multicolumn{1}{c|}{0}   & \multicolumn{1}{c|}{0}   & \multicolumn{1}{c|}{1}   & \multicolumn{1}{c|}{0}   & \multicolumn{1}{c|}{0}   & 1   \\ \cline{2-23} 
\multicolumn{1}{|c|}{}                                            & s20                                                  & 1                                                    & \multicolumn{1}{c|}{1}  & \multicolumn{1}{c|}{1}  & \multicolumn{1}{c|}{0}  & \multicolumn{1}{c|}{0}  & \multicolumn{1}{c|}{1}  & \multicolumn{1}{c|}{1}  & \multicolumn{1}{c|}{0}  & \multicolumn{1}{c|}{1}  & \multicolumn{1}{c|}{1}  & 0   & \multicolumn{1}{c|}{0}   & \multicolumn{1}{c|}{1}   & \multicolumn{1}{c|}{1}   & \multicolumn{1}{c|}{0}   & \multicolumn{1}{c|}{0}   & \multicolumn{1}{c|}{0}   & \multicolumn{1}{c|}{1}   & \multicolumn{1}{c|}{0}   & \multicolumn{1}{c|}{0}   & 1   \\ \cline{2-23} 
\multicolumn{1}{|c|}{}                                            & s21                                                  & 1                                                    & \multicolumn{1}{c|}{1}  & \multicolumn{1}{c|}{1}  & \multicolumn{1}{c|}{1}  & \multicolumn{1}{c|}{0}  & \multicolumn{1}{c|}{1}  & \multicolumn{1}{c|}{1}  & \multicolumn{1}{c|}{0}  & \multicolumn{1}{c|}{1}  & \multicolumn{1}{c|}{1}  & 0   & \multicolumn{1}{c|}{0}   & \multicolumn{1}{c|}{1}   & \multicolumn{1}{c|}{1}   & \multicolumn{1}{c|}{1}   & \multicolumn{1}{c|}{0}   & \multicolumn{1}{c|}{0}   & \multicolumn{1}{c|}{1}   & \multicolumn{1}{c|}{0}   & \multicolumn{1}{c|}{0}   & 1   \\ \cline{2-23} 
\multicolumn{1}{|c|}{}                                            & s22                                                  & 1                                                    & \multicolumn{1}{c|}{1}  & \multicolumn{1}{c|}{1}  & \multicolumn{1}{c|}{0}  & \multicolumn{1}{c|}{0}  & \multicolumn{1}{c|}{1}  & \multicolumn{1}{c|}{1}  & \multicolumn{1}{c|}{0}  & \multicolumn{1}{c|}{1}  & \multicolumn{1}{c|}{1}  & 0   & \multicolumn{1}{c|}{0}   & \multicolumn{1}{c|}{1}   & \multicolumn{1}{c|}{1}   & \multicolumn{1}{c|}{0}   & \multicolumn{1}{c|}{0}   & \multicolumn{1}{c|}{0}   & \multicolumn{1}{c|}{1}   & \multicolumn{1}{c|}{0}   & \multicolumn{1}{c|}{0}   & 0   \\ \cline{2-23} 
\multicolumn{1}{|c|}{}                                            & s23                                                  & 1                                                    & \multicolumn{1}{c|}{1}  & \multicolumn{1}{c|}{1}  & \multicolumn{1}{c|}{1}  & \multicolumn{1}{c|}{0}  & \multicolumn{1}{c|}{1}  & \multicolumn{1}{c|}{1}  & \multicolumn{1}{c|}{0}  & \multicolumn{1}{c|}{1}  & \multicolumn{1}{c|}{1}  & 0   & \multicolumn{1}{c|}{0}   & \multicolumn{1}{c|}{1}   & \multicolumn{1}{c|}{1}   & \multicolumn{1}{c|}{1}   & \multicolumn{1}{c|}{0}   & \multicolumn{1}{c|}{0}   & \multicolumn{1}{c|}{1}   & \multicolumn{1}{c|}{0}   & \multicolumn{1}{c|}{0}   & 1   \\ \cline{2-23} 
\multicolumn{1}{|c|}{}                                            & s24                                                  & 1                                                    & \multicolumn{1}{c|}{1}  & \multicolumn{1}{c|}{1}  & \multicolumn{1}{c|}{0}  & \multicolumn{1}{c|}{0}  & \multicolumn{1}{c|}{1}  & \multicolumn{1}{c|}{1}  & \multicolumn{1}{c|}{0}  & \multicolumn{1}{c|}{1}  & \multicolumn{1}{c|}{1}  & 0   & \multicolumn{1}{c|}{0}   & \multicolumn{1}{c|}{1}   & \multicolumn{1}{c|}{1}   & \multicolumn{1}{c|}{0}   & \multicolumn{1}{c|}{0}   & \multicolumn{1}{c|}{0}   & \multicolumn{1}{c|}{1}   & \multicolumn{1}{c|}{0}   & \multicolumn{1}{c|}{0}   & 0   \\ \cline{2-23} 
\multicolumn{1}{|c|}{}                                            & s25                                                  & 1                                                    & \multicolumn{1}{c|}{1}  & \multicolumn{1}{c|}{1}  & \multicolumn{1}{c|}{0}  & \multicolumn{1}{c|}{0}  & \multicolumn{1}{c|}{1}  & \multicolumn{1}{c|}{1}  & \multicolumn{1}{c|}{0}  & \multicolumn{1}{c|}{1}  & \multicolumn{1}{c|}{1}  & 0   & \multicolumn{1}{c|}{0}   & \multicolumn{1}{c|}{1}   & \multicolumn{1}{c|}{1}   & \multicolumn{1}{c|}{0}   & \multicolumn{1}{c|}{0}   & \multicolumn{1}{c|}{0}   & \multicolumn{1}{c|}{1}   & \multicolumn{1}{c|}{0}   & \multicolumn{1}{c|}{0}   & 0   \\ \cline{2-23} 
\multicolumn{1}{|c|}{}                                            & s26                                                  & 1                                                    & \multicolumn{1}{c|}{1}  & \multicolumn{1}{c|}{1}  & \multicolumn{1}{c|}{0}  & \multicolumn{1}{c|}{0}  & \multicolumn{1}{c|}{1}  & \multicolumn{1}{c|}{1}  & \multicolumn{1}{c|}{0}  & \multicolumn{1}{c|}{1}  & \multicolumn{1}{c|}{1}  & 0   & \multicolumn{1}{c|}{0}   & \multicolumn{1}{c|}{1}   & \multicolumn{1}{c|}{1}   & \multicolumn{1}{c|}{0}   & \multicolumn{1}{c|}{0}   & \multicolumn{1}{c|}{0}   & \multicolumn{1}{c|}{1}   & \multicolumn{1}{c|}{0}   & \multicolumn{1}{c|}{0}   & 0   \\ \cline{2-23} 
\multicolumn{1}{|c|}{}                                            & s27                                                  & 1                                                    & \multicolumn{1}{c|}{1}  & \multicolumn{1}{c|}{1}  & \multicolumn{1}{c|}{0}  & \multicolumn{1}{c|}{0}  & \multicolumn{1}{c|}{1}  & \multicolumn{1}{c|}{1}  & \multicolumn{1}{c|}{0}  & \multicolumn{1}{c|}{1}  & \multicolumn{1}{c|}{1}  & 0   & \multicolumn{1}{c|}{0}   & \multicolumn{1}{c|}{1}   & \multicolumn{1}{c|}{1}   & \multicolumn{1}{c|}{0}   & \multicolumn{1}{c|}{0}   & \multicolumn{1}{c|}{0}   & \multicolumn{1}{c|}{1}   & \multicolumn{1}{c|}{0}   & \multicolumn{1}{c|}{0}   & 0   \\ \cline{2-23} 
\multicolumn{1}{|c|}{\multirow{-10}{*}{m6}}                       & s28                                                  & 1                                                    & \multicolumn{1}{c|}{1}  & \multicolumn{1}{c|}{1}  & \multicolumn{1}{c|}{0}  & \multicolumn{1}{c|}{0}  & \multicolumn{1}{c|}{1}  & \multicolumn{1}{c|}{1}  & \multicolumn{1}{c|}{0}  & \multicolumn{1}{c|}{1}  & \multicolumn{1}{c|}{1}  & 0   & \multicolumn{1}{c|}{0}   & \multicolumn{1}{c|}{1}   & \multicolumn{1}{c|}{1}   & \multicolumn{1}{c|}{0}   & \multicolumn{1}{c|}{0}   & \multicolumn{1}{c|}{0}   & \multicolumn{1}{c|}{1}   & \multicolumn{1}{c|}{0}   & \multicolumn{1}{c|}{0}   & 0   \\ \hline
\multicolumn{2}{|c|}{Test Outcome}                                                                                       & F                                                    & \multicolumn{1}{c|}{P}  & \multicolumn{1}{c|}{P}  & \multicolumn{1}{c|}{P}  & \multicolumn{1}{c|}{P}  & \multicolumn{1}{c|}{P}  & \multicolumn{1}{c|}{P}  & \multicolumn{1}{c|}{P}  & \multicolumn{1}{c|}{P}  & \multicolumn{1}{c|}{P}  & P   & \multicolumn{1}{c|}{P}   & \multicolumn{1}{c|}{P}   & \multicolumn{1}{c|}{P}   & \multicolumn{1}{c|}{P}   & \multicolumn{1}{c|}{P}   & \multicolumn{1}{c|}{P}   & \multicolumn{1}{c|}{P}   & \multicolumn{1}{c|}{P}   & \multicolumn{1}{c|}{P}   & P   \\ \hline
\end{tabular}
}
\vspace{-0.1cm}
\label{tab:cov}
\end{table}

\begin{table}[htp]
\centering
\caption{Suspicious values for the suspicious methods with ten tests selected by RLFDC}
\vspace{-0.25cm}
\begin{tabular}{c|c|c|c|c|c|c|c|c|c|c|c}
\hline
   & t0      & t1      & t2      & t3      & t4      & t5      & t6      & t7      & t8      & t9      & t10     \\ \hline
m1 & 1.000/6 & 0.707/6 & 0.577/6 & 0.500/6 & 0.447/6 & 0.408/6 & 0.378/6 & 0.354/6 & 0.333/6 & 0.316/6 & 0.316/6 \\ \hline
m2 & 1.000/6 & 0.707/6 & 0.707/3 & 0.707/3 & 0.577/4 & 0.577/3 & 0.577/3 & 0.577/2 & 0.577/2 & 0.577/2 & 0.577/2 \\ \hline
m3 & 1.000/6 & 0.707/6 & 0.707/3 & 0.707/3 & 0.577/4 & 0.577/3 & 0.577/3 & 0.500/3 & 0.500/3 & 0.500/3 & 0.500/3 \\ \hline
\textcolor{red}{m4} & 1.000/6 & 0.707/6 & 0.707/3 & 0.707/3 & 0.707/1 & 0.707/1 & 0.707/1 & 0.707/1 & 0.707/1 & 0.707/1 & 0.707/1 \\ \hline
m5 & 1.000/6 & 0.707/6 & 0.577/6 & 0.500/6 & 0.500/5 & 0.447/5 & 0.408/5 & 0.408/5 & 0.408/5 & 0.408/4 & 0.408/4 \\ \hline
m6 & 1.000/6 & 0.707/6 & 0.577/6 & 0.577/4 & 0.577/4 & 0.500/4 & 0.447/4 & 0.447/4 & 0.408/5 & 0.378/5 & 0.378/5 \\ \hline
\end{tabular}
\vspace{-0.1cm}
\label{tab:value-rlfdc}
\end{table}

\begin{table}[htp]
\centering
\caption{Suspicious values for the suspicious methods with ten tests selected by TfD}
\vspace{-0.25cm}
\begin{tabular}{c|c|c|c|c|c|c|c|c|c|c|c}
\hline
   & t0      & t11     & t12     & t13     & t14     & t15     & t16     & t17     & t18     & t19     & t20     \\ \hline
m1 & 1.000/6 & 0.707/6 & 0.577/6 & 0.500/6 & 0.447/6 & 0.447/6 & 0.447/6 & 0.408/6 & 0.408/6 & 0.408/6 & 0.378/6 \\ \hline
m2 & 1.000/6 & 1.000/5 & 1.000/3 & 1.000/3 & 1.000/3 & 1.000/3 & 1.000/3 & 1.000/3 & 1.000/3 & 1.000/3 & 1.000/3 \\ \hline
m3 & 1.000/6 & 1.000/5 & 1.000/3 & 1.000/3 & 1.000/3 & 1.000/3 & 1.000/3 & 1.000/3 & 1.000/3 & 1.000/3 & 1.000/3 \\ \hline
\textcolor{red}{m4} & 1.000/6 & 1.000/5 & 1.000/3 & 1.000/3 & 1.000/3 & 1.000/3 & 1.000/3 & 1.000/3 & 1.000/3 & 1.000/3 & 1.000/3 \\ \hline
m5 & 1.000/6 & 1.000/5 & 0.707/5 & 0.707/4 & 0.707/4 & 0.707/4 & 0.707/4 & 0.707/4 & 0.707/4 & 0.707/4 & 0.577/4 \\ \hline
m6 & 1.000/6 & 1.000/5 & 0.707/5 & 0.577/5 & 0.577/5 & 0.577/5 & 0.577/5 & 0.500/5 & 0.500/5 & 0.500/5 & 0.500/5 \\ \hline
\end{tabular}
\label{tab:value-tfd}
\end{table}

We use the following motivating example to illustrate how an FDC metric facilitates test selection for FL. This example, derived from the \textit{Math} project within Defects4J~\cite{just2014defects4j}, represents a simplified real-world example. There is an initial failing test $t0$ which reveals a fault in the project. Our aim is to select additional tests with FDC metrics to help developers pinpoint the buggy method. Table~\ref{tab:cov} shows the coverage information and test outcomes for the ten tests selected by RLFDC and TfD, respectively (1/0 demonstrates cover/uncover and F/P demonstrates Fail/Pass). For simplicity, we only show the coverage matrix for the 26 statements, spanning six methods, that are covered by the failing test $t0$. The buggy statements and methods are highlighted in red. Table~\ref{tab:value-rlfdc} and Table~\ref{tab:value-tfd} show the change of ``suspicious score/method ranking'' of each suspicious method when additional tests are selected by RLFDC and TfD, respectively. The suspicious scores are computed with a widely-studied FL technique Ochiai~\cite{ochiai} using the max-tie-breaker~\cite{an2022fdg}. We aggregate the line-level Ochiai scores at the method level with the highest score of all the lines in the method~\cite{sohn2017fluccs}.

From Table~\ref{tab:value-rlfdc} and Table~\ref{tab:value-tfd}, we find that with ten tests selected by RLFDC, the buggy method $m4$ is ranked at the 1st, while with ten tests selected by TfD, it is ranked at the 3rd, which demonstrates the superiority of RLFDC. We further explain how this result comes.
Initially, all the suspicious statements are covered by the initial failing test $t0$. Therefore, all the suspicious methods have the same suspicious score and cannot be discriminated from one another. With the max-tie-breaker strategy, they are all ranked at the 6th. For RLFDC, it considers two features to value each test, \textit{cover} and \textit{split} (detailed in Section~\ref{sec:model_design}), and can adaptively balance the weight between the two features based on the current state thanks to reinforcement learning. For the \textit{cover} feature, it measures how the additional test covers the suspicious statements since a passing test covering more statements can help reduce the suspiciousness of more non-buggy statements and help pinpoint the fault faster. 
For the \textit{split} feature, it measures how an additional test distinguishes statements that were previously indistinguishable. Specifically, statements covered by the same set of tests are referred to as an ``ambiguity group'' in Section~\ref{sec:model_design}. Within an ambiguity group, all statements share the same program spectrum since they are executed by the same set of tests. Consequently, these statements are assigned the same suspicious value when spectrum-based fault localization techniques are applied, resulting in an indistinguishability problem. To address this, RLFDC incorporates the \textit{split} feature, aiming to break the ambiguity groups formed by existing tests.
In the example, RLFDC first selects tests that cover suspicious statements as many as possible. Therefore, it selects $t1$ and $t2$ which cover 28 and 18 statements respectively, and the buggy method $m4$ is ranked 3rd with the two tests. Note that $m4$ now has the same score with $m2$ and $m3$, because the buggy statement $s14$ now is in the same ambiguity group with $s2$, $s3-s8$ and they cannot be discriminated from each other. Then RLFDC finds that with $t1$ and $t2$, there are many ambiguity groups formed (e.g., $s2-s9$, $s10-s13$). Based on the current state, RLFDC turns its attention to splitting those ambiguity groups to discriminate the statements and help more accurately pinpoint the fault. In particular, it selects $t3$ and $t4$ to break the groups. The test $t4$ covers $s2$, $s3-s8$ but does not cover $s14$. Therefore, the group is broken and $s14$ has the highest suspicious score, making $m4$ ranked at the 1st. As for the state-of-the-art result-agnostic metric TfD, it only considers the number of ambiguity groups. Therefore, it aims at discriminating statements as much as possible by selecting $t11$ and $t12$. However, there are many ambiguity groups and it tries to split the trivial groups in those non-buggy methods $m1$, $m5$, and $m6$, failing to discriminate statements in $m2$, $m3$, and $m4$. Therefore, the buggy method $m4$ is finally ranked at the 3rd. Note that compared to TfD, RLFDC splits ambiguity groups in a more effective way since RLFDC also considers the \textit{cover} feature. Therefore, RLFDC will select $t11$ with a smaller priority since $t11$ can only discriminate one statement $s1$.

Despite incorporating the two features, RLFDC utilizes reinforcement learning (RL) to automatically and adaptively determine their optimal balance based on the current state, including the number of tests and ambiguity groups. Specifically, RLFDC improves its accuracy by utilizing fault localization feedback as the RL reward, thereby achieving a more precise FDC measurement. This precision stems from the fact that the fault localization contribution directly reflects the FDC value of a test. We will provide a detailed explanation of our RL framework in the following section.

\section{Methodology}

We propose a RL-based approach to automatically construct an FDC metric, i.e., \textbf{R}eforcement \textbf{L}earning based \textbf{F}ault \textbf{D}iagnosis \textbf{C}apability (\techname{}). Unlike existing FDC metrics designed based on pre-defined heuristics, \techname{} is constructed systematically by learning the measuring strategy based on direct FL feedback. The key advantage of RL is that RL uses reward to update the model while traditional supervised learning approaches need labels to train the model. However, in such a scenario, it is hard to get labels for the FDC values of tests because FDC is influenced not only by the tests themselves but also by the underlying test suite, as demonstrated in the motivating example. Therefore, supervised learning is inapplicable and thus we use direct FL feedback to calculate reward to train our model.

\begin{figure}
    \centering
    \includegraphics[width=0.8\linewidth]{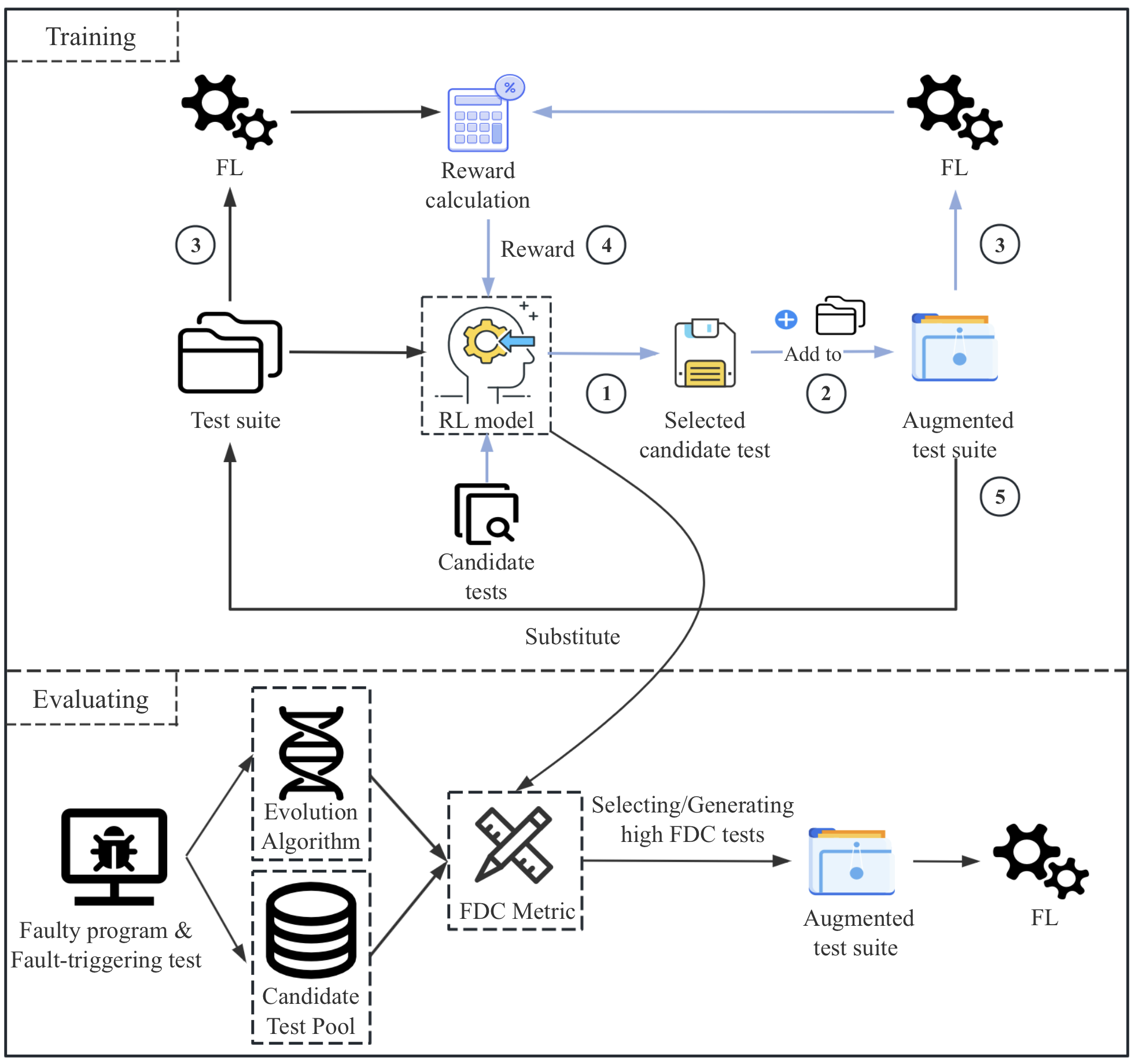}
    \caption{Boosting FL with \techname{}}
    \label{fig:tech_overview}
\end{figure}

Our approach consists of two main stages, i.e., a training stage and an evaluating stage, as shown in Fig.~\ref{fig:tech_overview}. 
In the training stage, an RL model predicts an FDC value of each candidate test, based on which it sequentially selects individual tests and continuously improves based on the FL feedback with the selected tests. 
The training stage contains a number of iterations, each containing five steps (i.e., \ding{172}-\ding{176} in Fig.~\ref{fig:tech_overview}., detailed in Section~\ref{sec:training_process}). At the end of the training stage, a predictive FDC model is built. 
In the evaluating stage, the trained RL model is leveraged as an FDC metric (\techname{}) to predict the FDC values of candidate tests to boost FL with test selection or generation.
In the following, 
Section~\ref{sec:model_design} presents the RL design of our RL-based approach. Section~\ref{sec:training_process} presents the training stage of it. Section~\ref{sec:evaluating_process} presents the evaluating stage of it.

\subsection{RL Design}
\label{sec:model_design}
We first introduce the RL design of our approach. RL is a machine learning framework, which aims to learn a strategy to maximize the \textbf{reward}. It involves the interaction of an agent and an environment, through which the agent finds an optimal strategy to maximize the expected reward. The agent performs a series of \textbf{actions} based on a series of \textbf{states} of the environment. The environment measures the influence of the series of actions and returns a series of \textbf{reward}, which helps the agent continuously adjust its strategy. 

In this paper, we build a better FDC metric by modeling the test selection process for FL as an RL process. Initially, there is one test suite $T$ containing only one failing test and a set of candidate tests to be selected to boost the FL performance. During the test selection process, the RL model plays the role of ``agent''. The coverage information of the whole test suite $T$ (including the original failing test and selected tests) forms the ``state''. The selection of tests forms the ``action'', while the results produced by an FL technique establish the basis of ``reward''. To maximize the reward, the RL model predicts the value of each ``action'' (i.e., the FDC value of each test), and then selects the ``action'' with the highest value. Based on the reward (i.e., the influence of the selected tests on the FL results), the model updates its strategy on value prediction (i.e., improving the accuracy of \techname{} on FDC measurement).



\subsubsection{RL Model Design}
\label{sec:rl_model_design}
We utilize a deep neural network as our RL model, leveraging its strong representation power and outstanding effectiveness~\cite{li2019deepfl,lou2021boosting}. The input of the network consists of state-action pairs, and the output of the network is the value of each action taken in the current state (i.e., the FDC value of each test considering the underlying test suite $T$). The input is extracted using coverage information of the test suite and the candidate tests. In particular, the state depends on the test suite $T$, while the action depends on the test suite and the candidate tests. More design details on the state, action, and reward of our RL-based approach are given in Section~\ref{sec:design}. The state is first embedded through an embedding layer consisting of two linear projection layers and two $\operatorname{ReLU}$ activation layers as follows.
\begin{equation}
    \label{equ:cal_state}
    \phi(state) = \operatorname{ReLU}(\operatorname{Linear}(\operatorname{ReLU}(\operatorname{Linear}(state))))
\end{equation}
The embedded state is then concatenated with the action vector and is fed into our FDC prediction network. 

The FDC prediction network consists of three linear layers. Formally, the output of each layer is computed by
\begin{equation}
    \label{equ:cal_FDC}
     \bm y^{(l)}= W^{(l)} \bm y^{(l-1)} + \bm b^{(l)}, \quad l \in 
     \{1,2,3\}
\end{equation}
where $l$ denotes the $l$-th layer, $\bm y^{(l)}$ denotes the output of the $l$-th layer, $\bm y^{(0)}=[\phi(state); action]$, $W^{(l)}$ and $\bm b^{(l)}$ denote trainable variables. In particular, we use the activation function $\operatorname{ReLU}$ for the first two layers. We use the output of the last layer, which is a singular value, as the predicted FDC value, i.e., $\mathrm{FDC} = y^{(3)}$. The FDC value represents the potential gain of selecting or generating a test for FL, which is further used for test selection and generation in the evaluating stage.

\subsubsection{State \& Action \& Reward Design}
\label{sec:design}

In this section, we introduce the state, action, and reward design of our RL-based approach. Let us assume there is an $m \times n$ coverage matrix $C$ denoting the relation between a test suite $T$ and code elements $E$, where $C[i,j] = 1$ denotes test $t_i$ covers element $e_j$ and $C[i,j]=0$ otherwise. Given a test $t_i$, we have its coverage vector $ct_i$, where $ct_i$ is the $i$th row of $C$. Given a code element $e_j$, we have its execution vector $ce_j$, where $ce_j$ is the $j$th column of $C$. $AG(T)$ denotes the set of ambiguity groups under test suite $T$. An ambiguity group $ag \in AG(T)$ is a set of code elements that share the same execution vector and thus cannot be distinguished from each other.

\textit{(1) State.} The state is defined based on the current test suite $T$. The current test suite is an indispensable factor that should be considered since the FDC value of a test depends not only on the test itself but also on the test suite where it is supposed to be added. We use two features to describe the state space: the number of tests in the test suite (denoted as $num\_tests$) and the number of ambiguity groups in the coverage matrix of the test suite (denoted as $num\_ag$), i.e., a state can be represented as:
\begin{equation}
    \label{equ:rep_state}
    state=(num\_tests, num\_ag)
\end{equation}
The number of tests demonstrates the scale of the current test suite. The number of ambiguity groups demonstrates to what extent the current test suite distinguishes the code elements in FL. In other words, the two features describe how much effort has been made (i.e., the number of selected tests) and how many difficulties still exist (i.e., the number of ambiguity groups).

\textit{(2) Action.} The action is defined based on candidate tests. We aim at selecting or generating tests to diagnose faults. From one aspect, tests that split suspicious ambiguity groups, i.e., tests executing different subsets of suspicious ambiguity groups, can help distinguish the elements and thus help diagnose faults. However, solely focusing on splitting ambiguity groups may fall into local optimums. A passing test simultaneously covering several ambiguity groups rather than splitting them may also help pinpoint the fault, because it reduces the suspiciousness of the non-buggy elements in these groups. 
Building on this intuition and the inspiration for combining the two aspects (i.e., respectively termed as ``Cover'' and ``Split'') introduced in previous work~\cite{an2022fdg}, we also use two features, \textbf{cover} and \textbf{split}, but adapt them in a result-agnostic way, to describe the potential value of a candidate test $t$ when added to the underlying test suite $T$.
Thus, the value of a test can be represented by its ability to \textbf{cover} suspicious elements and the ability to \textbf{split} ambiguity groups:

\begin{equation}
    \label{equ:rep_action}
    action=(cover, split)
\end{equation}

For the \textit{cover} feature, we adapt the prior result-aware metric Prox~\cite{an2022fdg} by ignoring the execution results of all tests except the given failing test\footnote{Prox calculates the average Jaccard distance between a candidate test and each failing test, as Prox considers the execution results of all tests as input.}. The \textit{cover} feature is defined as follows.
\begin{equation}
    \label{equ:cal_cover}
    cover(T,t) = \operatorname{Jaccard}\left(ct_{t}, ct_{fail}\right)
\end{equation}
where $\operatorname{Jaccard}$ denotes the Jaccard distance between two coverage vectors, $ct_{t}$ and $ct_{fail}$ denote the coverage vector of the candidate test $t$ and the given failing test respectively. Therefore, $cover$ assigns higher priority to tests that can cover more suspicious elements.

For the \textit{split} feature, inspired by Hao et al.~\cite{hao2010test}\footnote{Although both works (i.e., prior work~\cite{hao2010test} and this work) are defined based on the same intuition, the prior work uses execution results of all tests and a different definition of ambiguity group, which makes the two works different.}, we measure the ability of a test to distinguish suspicious elements based on: (1) whether this test splits a large ambiguity group containing many suspicious elements and 
(2) to what extent the split made by this test helps reduce the number of suspicious elements. 
To measure the former, for an ambiguity group $ag$, we define $priority(ag)$ as the number of code elements in $ag$. Intuitively, a large ambiguity group tends to have a higher probability of containing a buggy element and thus deserves more attention. To measure the latter, for an ambiguity group $ag$, we define $div(t,ag)$ as the impact of the test $t$ on the ambiguity group $ag$. As evenly splitting an ambiguity group tends to result in better FL performance~\cite{hao2010test}, we calculate $div(t,ag)$ based on the size of the smaller subset when $t$ is used to divide $ag$ to two subsets based on coverage, i.e., the minimum between $\left|\left\{e_j \in ag \mid ct_{t}[j]=1\right\}\right|$ and $\left|\left\{e_j \in ag \mid ct_{t}[j]=0\right\}\right|$, where $ct_t$ denotes the coverage vector of the test $t$. Therefore, $div(t,ag)$ reaches the largest value when the group $ag$ is split evenly. Based on $priority(ag)$ and $div(t,ag)$, the \textit{split} feature is defined as follows. 
\begin{equation}
    \label{equ:cal_split}
    split(T,t) = \sum_{ag \in A G(T \cup t)} priority(ag) \cdot div(t, ag)
\end{equation}
where $AG(T \cup t)$ denotes the set of ambiguity groups formed if the candidate test $t$ is added to the test suite $T$.

\textit{(3) Reward.} We compute the reward based on the impact of the test $t$ on the performance of Spectrum-Based Fault Localization (SBFL). An SBFL technique takes the coverage matrix $C$ and test execution results as input and outputs a ranked list of program elements based on the descending order of their suspicious values. The earlier the list ranks the buggy element, the better performance the corresponding SBFL technique achieves. Following this intuition, we use the rank improvement of the buggy element resulting from the selected test $t$ as the reward, which measures the contribution of the test $t$ to fault diagnosis as follows. 
\begin{equation}
    \label{equ:cal_reward}
    reward(T,t) = ( rank(FL^{T_{init}}) - rank(FL^{T \cup t})) / rank(FL^{T_{init}})
\end{equation}
where $rank(FL^{T_{init}})$ denotes the highest rank of the buggy elements\footnote{Since there may exist multiple buggy elements in a program, we calculate the ranks of all buggy elements and choose the highest one.} using an SBFL technique with the initial test suite $T_{init}$, $rank(FL^{T \cup t})$ denotes the result after adding test $t$ to the current test suite $T$. We divide the difference by $rank(FL^{T_{init}})$ to get the relative changed rank. Note that we do not use $rank(FL^{T}) - rank(FL^{T \cup t})$ to calculate the difference because some temporarily ``bad'' performance test (i.e., resulting in $rank(FL^{T})< rank(FL^{T \cup t})$) may later benefit FL. The reward is only calculated in the training stage, but is not calculated in the evaluating stage. 


\subsection{Training Stage}
\label{sec:training_process}

In this section, we present the training stage of our approach. As shown by Fig.~\ref{fig:tech_overview}, the training stage begins with a given test suite $T$ (which contains only a given failing test initially) and a pool of candidate tests $T_c$. The training stage aims to build an FDC metric \techname{} for later usage (e.g., for a project $P$). Thus, this process is conducted on some historical versions of $P$ or on other projects. 

The training stage contains a number of training iterations. 
Each training iteration is a test selection process for FL where the RL model plays the role of selection criterion and the FL feedback is used to improve the accuracy of the RL model. In particular, for each training iteration, we repeat five steps: \ding{172} The RL model predicts the FDC value of each candidate test in $T_c$ and selects the candidate test with the largest FDC value. \ding{173} The selected candidate test $t_{sel}$ is added to $T$, and removed from $T_c$. \ding{174} An SBFL technique is applied to $T_{init}$ and $T \cup t_{sel}$, and then the difference between the rank lists is calculated, which is utilized as the reward of RL. \ding{175} The reward is forwarded to the RL model to notify the actual FDC value of the selected test $t_{sel}$. The RL model uses this FL feedback to adjust its prediction strategy. \ding{176} The augmented $T$ (i.e., $T \cup t_{sel}$) is utilized to substitute the previous $T$, which ends this iteration. 

More specifically, we use double Q-learning with experience replay~\cite{hasselt2010double} to train our RL model (i.e., Q network in Algorithm~\ref{alg_doubleq}). As Q-learning performs poorly due to large overestimation of action values~\cite{hasselt2010double}, we adopt double Q-learning, which uses an extra target Q network to approximate the maximum expected value. We select double Q-learning as it’s more suitable for high-accuracy regime tasks~\cite{xiong2020finite,van2016deep,hasselt2010double} and it is one of the most successful value-based, model-free RL algorithms, which suits our scenario where the network needs to predict FDC values (i.e., value-based) and has no prior-knowledge of the environment (i.e., model-free).

\begin{algorithm}[H]
\renewcommand{\algorithmicrequire}{\textbf{Input:}}
\renewcommand{\algorithmicensure}{\textbf{Output:}}
\caption{Double Q-learning with experience replay}
\label{alg_doubleq}
\setlength{\intextsep}{0.1cm}
    \begin{algorithmic}[1]
        \Require initial failing test $t_{fail}$, candidate test set $T_{c}$, capacity $N$, step $K$, episode $M$, updating step $C$, learning step $L$, discount factor $\gamma$, random selection probability $\sigma$.
        \Ensure A trained Q network.
        
        \State Initialize replay memory $D$ to capacity $N$
        \State Initialize step counter $counter$ to 0
        \State Initialize $Q$ network with random weights $\theta$
        \State Initialize target $\hat{Q}$ network with weights $\theta^{-} = \theta$
        \For{$episode=1$ to $M$}
            \State Initialize test suite $T = \{t_{fail}\}$
            \State initBuggyRank = doFaultLocalization($T$)
        
            \For{$k=1$ to $K$}
                \State $s_k$ = getState$(T)$
                \For{each candidate test $t_i \in T_{c}$}
                    \State $a_{ki}$ = getAction$(T, t_i)$
                    \State $fitness_{ki} = Q(\phi(s_k), a_{ki};\theta)$
                \EndFor
                \State Select a random test $t_{sel} \in T_{c}$ with a probability $\sigma$, otherwise select $t_{sel}$ who has the highest $fitness_k$
                \State $T_{c}$=$T_{c} \setminus \{t_{sel}\}$ and $T=T \cup t_{sel}$
                \State curBuggyRank = doFaultLocalization($T$)
                \State $r_k$ = reward(initBuggyRank, curBuggyRank)
                \State $s_{k+1}$ = getState$(T)$
                \State Store transition $(s_k,a_k,r_k,s_{k+1})$ in $D$
                \State $counter$ = $counter + 1$
                \If{($|D|=N $) and ($counter$ $\operatorname{mod} L == 0$)}
                    \State Sample random minibatch of transitions $(s_j,a_j,r_j,s_{j+1})$ from $D$
                    \If{$j+1=K$} 
                        \State $y_j=r_j$
                    \Else
                        \State $y_j=r_j+\gamma \max _{a^{\prime}} \hat{Q}\left(\phi(s_{j+1}), a^{\prime} ; \theta^{-}\right)$
                    \EndIf
                    \State Perform a gradient descent step on $\left(y_j-Q\left(\phi(s_j), a_j ; \theta\right)\right)^2$ with respect to the network parameters $\theta$
                    \State Every $C$ steps reset $\hat{Q} = Q$
                \EndIf
            \EndFor
        \EndFor
    \end{algorithmic}  
\end{algorithm}

Algorithm~\ref{alg_doubleq} formally presents the double Q-learning algorithm used in our approach. It takes the initial failing test $t_{fail}$, the candidate test set $T_{c}$, a series of parameters as input, and returns the trained Q network model (i.e., the trained RL model). First, it initializes a memory pool $D$, step counter $counter$, network $Q$, and target network $\hat{Q}$ using the input parameters (Lines 1-4). Then, it utilizes $M$ episodes to train the Q network (Lines 5-32). In each episode, it first initializes the test suite $T$ using $t_{fail}$, then calculates the initial buggy element rank with an FL technique (Lines 6-7). The algorithm uses $K$ steps (i.e., iterations) to complete an episode, where in each step, a candidate test is selected, and the transition is stored (Lines 8-31). For each step, the state for the test suite and the action for each candidate test can be calculated using Eq.~\ref{equ:rep_state} and Eq.~\ref{equ:rep_action}, then the fitness for each candidate test can be calculated using Eq.~\ref{equ:cal_FDC} (Lines 9-13). To ensure that the model does exploration rather than greedily falls into local optimums, the test $t_{sel}$ is randomly selected with a probability of $\sigma$ and is selected based on the fitness otherwise (Line 14). The selected test $t_{sel}$ is removed from the candidate test set and added to the test suite $T$ (Line 15). Then we apply the FL technique with the augmented test suite to get the current buggy element rank, and calculate the reward according to Eq.~\ref{equ:cal_reward} (Lines 16-17). The transition for this step is stored in $D$, and a counter records the number of steps (Lines 18-20). When the replay memory reaches its capacity, for every $L$ steps, we sample a random minibatch from $D$ and update the parameters accordingly (Lines 21-30). The expected reward $y$ is calculated using the target network $\hat{Q}$ (Lines 23-27). We use mean-squared error to calculate the difference between the predicted value and the expected reward $y$ (Line 28). The parameters of $\hat{Q}$ are fixed and updated every $C$ steps by being substituted with parameters from $Q$ (Line 29).


\subsection{Evaluating Stage}
\label{sec:evaluating_process}

As shown in Fig.~\ref{fig:tech_overview}, the evaluating stage presents the usage of \techname{}, which is the trained RL model resulting from the training stage. In particular, in the evaluating stage, we apply \techname{} to two FL-oriented scenarios, i.e., test selection and generation for FL. \textbf{Note that in the evaluating stage, we do not feed any execution results of tests as input to the RL model} (FL results are only utilized in the reward calculation, which is only needed in the training stage). Thus, the trained RL model, i.e., \techname{}, is a result-agnostic metric.

Given a fault-triggering test and a faulty program, test selection for FL aims to select some tests from a given candidate test pool (without oracles) and use them to augment the given test suite for FL. In this scenario, we use \techname{} to guide test selection, i.e., iteratively selecting a test with the largest FDC value given by \techname{} and adding the selected test to augment the test suite, which initially contains only the given fault-triggering test. The process is repeated until some termination condition is satisfied, e.g., achieving some FL performance goals or reaching some resource limits. 



Given a fault-triggering test and a faulty program, test generation for FL aims to generate more tests for FL automatically, and thus \techname{} can serve as a fitness function in search-based test generation. For example, EVOSUITE is a test generation tool based on evolutionary algorithm, and we integrate \techname{} with it to facilitate test generation for FL. In particular, in the evolutionary algorithm, tests with the largest FDC value predicted by \techname{} are selected, and are used as parents to produce offsprings (i.e., more tests).


\section{Evaluation setup}


In this section, we evaluate \techname{} in FL-oriented test case selection and generation tasks, as prior work does~\cite{ddu,tfd,entbug,an2022fdg}. 
In particular, in the former task, we use \techname{} and other metrics as selection criteria to select tests and evaluate the FL performance with the corresponding selected test cases. In the latter task, we use \techname{} and other metrics as fitness functions to guide test generation and evaluate the FL performance with the corresponding generated tests. To sum up, we aim to answer five research questions (RQs).


RQ1. \textbf{How does \techname{} perform on human-written test selection?} To answer this RQ, we select tests based on human-written tests using different FDC metrics, and compare the FL performance of the various augmented test suites. This serves as an ideal evaluation setting, featuring tests written by experienced developers with deep knowledge of the projects~\cite{an2022fdg}.

RQ2. \textbf{How does \techname{} perform on automatically generated test selection?} Different from RQ1, in this RQ we select tests from a test pool automatically generated by a search-based test generation tool EVOSUITE. Besides, to further investigate the generalizability of FDC metrics, we also evaluate the test selection task on a test pool automatically generated by Randoop~\cite{pacheco2007randoop}.


RQ3. \textbf{How does \techname{} perform on automated test generation?} To answer this RQ, we utilize various metrics (including \techname{}) as fitness functions in EVOSUITE, and compare the FL performance of their generated tests.

RQ4. \textbf{How does \techname{} perform in cross-project scenario?} The proposed \techname{} is built based on a training process that simulates test selection for FL. This process can be conducted on cross-version scenario and cross-project scenario. The evaluation for RQ1-RQ3 is conducted in the former scenario, whereas in this RQ, we investigate the performance of \techname{} in the latter scenario.

RQ5. \textbf{How do different components of \techname{} contribute to its effectiveness?} We perform an ablation study to validate the design decisions of our approach. To answer this RQ, we investigate the impact of the designed features, the network structure, and the training algorithm to the effectiveness of \techname{} by conducting experiments on different variants of \techname{}.



\begin{table}[!htp]
    \vspace{-0.1cm}
    \setlength{\tabcolsep}{2pt}
    \centering
    \caption{Subjects}
    \vspace{-0.2cm}
    \begin{tabular}{l|c|c|c|c|c|c}
    \toprule \multirow{2}{*}{ Subject } & \multirow{2}{*}{ \# Faults } & \multirow{2}{*}{kLoC} & \multicolumn{2}{|c|}{\text { Avg. \# Tests }} & \multicolumn{2}{|c}{\text { Avg. \# Methods }} \\
    \cmidrule{4-7} & & & \text { Total } & \text { Failing } & \text { Total } & \text { Buggy } \\
    \midrule \text { Commons-\textbf{lang} } & 62 & 22 & 1,819 & 2.0 & 2,180 & 1.5 \\
     \text { Commons-\textbf{math} } & 106 & 85 & 2,524 & 1.7 & 4,544 & 1.7 \\
     \text { JFree\textbf{Chart} } & 26 & 96 & 1,814 & 3.5 & 7,478 & 4.5 \\
     \text { Joda-\textbf{Time} } & 26 & 28 & 3,918 & 2.8 & 3,804 & 2.0 \\
     \text { \textbf{Closure} compiler } & 131 & 90 & 7,211 & 2.6 & 8,203 & 1.7 \\
    \midrule \text { Total } & 351 & \multicolumn{5}{c}{} \\
    \bottomrule
    \end{tabular}
    \vspace{-0.2cm}
    \label{tab:subject}
\end{table}

\subsection{Subjects}
In this study, following prior work~\cite{an2022fdg}, we use five projects from Defects4J (V2.0.0)~\cite{just2014defects4j}, a Java benchmark widely used in FL~\cite{li2019deepfl,lou2020can,lou2021boosting,zou2019empirical}. Each project has some buggy versions, while each buggy version contains one real fault in the program and at least one test revealing the existence of this fault. All the tests provided by Defects4J are manually constructed. 
We remove six deprecated faults from our study because they are irreproducible due to behavioral changes introduced under Java 8~\cite{d4jwebsite2024}.
Table~\ref{tab:subject} presents the statistics of the projects used in this study. Column ``Subject'' presents the full name of the subject. Column ``\# Faults'' presents the number of buggy versions for each subject. Columns ``Total'' and ``Failing'' under ``Avg. \# Tests'' present the average number of total tests and failing tests for each buggy version. Columns ``Total'' and ``Buggy'' under ``Avg. \# Methods'' present the average number of total methods and buggy methods for each buggy version. We use the abbreviation of project names (stressed in bold font) in the remainder of this paper. Note that in RQ2 and RQ3, we remove another 103 faults because EVOSUITE cannot generate compilable tests for the corresponding buggy programs.


\subsection{Baselines}

In the literature, the metrics to measure FDC are classified into result-aware and result-agnostic metrics based on the utilization of test results~\cite{an2022fdg}. To conduct a comprehensive comparison, we include three state-of-art result-agnostic metrics EntBug~\cite{entbug}, DDU~\cite{ddu} and TfD~\cite{tfd}, and one state-of-the-art result-aware metric FDG~\cite{an2022fdg} as our baselines. Note that our \techname{} is a result-agnostic metric.

FDG~\cite{an2022fdg} is a recently proposed result-aware metric that employs SBFL scores to learn which part of the program requires additional diagnostic information. It is shown to perform the best on the test selection task for FL~\cite{an2022fdg}. 
FDG uses the SBFL scores to assign weights to ambiguity groups and program elements, where the execution result of the newly-added test is needed. $\text{FDG}(T,t)=\alpha \cdot (1-\frac{1}{n-1} \cdot \sum_{ag \in A G(T \cup\{t\})} p(ag) \cdot(|ag|-1)) + (1-\alpha) \cdot \frac{\sum_{j=1}^n w_j \cdot ct_t[j]}{n}$, where $w_j$ is the SBFL score of $e_j$, $ct_t$ is the coverage vector of test $t$, $n$ is the number of code elements, $\alpha$ is a hyperparameter that needs to be tuned, $p(ag)=\sum_{e_j \in ag} p\left(e_j\right)$, $p\left(e_j\right)=w_j / \sum_i w_i$.

EntBug~\cite{entbug} is a result-agnostic metric based on entropy. In particular, EntBug applies probability theory concepts to minimize the uncertainty in the diagnostic ranking, and is defined using the density of the coverage matrix. $\text{EntBug}(T)=1-|1-2\cdot\rho(T)|$, where $\rho(T)=\sum_{i=1}^m \sum_{j=1}^n C[i,j] /(m \times n)$. Recall that coverage matrix $C$ has the shape of $ m\times n$ and $C[i,j]=1/0$ denotes whether test $t_i$ covers/uncovers program element $e_j$. 

TfD~\cite{tfd} is a result-agnostic metric based on the number of ambiguity groups in the test suite $T$, i.e., $\text{TfD}=|AG(T)|$. It is designed with the intuition that the more ambiguity groups a coverage matrix has, the less ambiguity the program spectrum has.

DDU~\cite{ddu} is a combined, result-agnostic metric. $\text{DDU}(T)=density(T)\times diversity(T) \times uniqueness(T)$, where $density(T)$ is equal to $\rho(T)$, $uniqueness(T)$ is equal to $|AG(T)|/n$, $diversity(T)$ is the Gini-Simpson index~\cite{jost2006entropy} among the rows of the coverage matrix. DDU is proposed to combine these three key properties of a coverage matrix for achieving more accurate FL.

In summary, these metrics are all hand-crafted formulas designed based on some heuristics, which leaves room for harnessing the power of machine learning to learn a more accurate metric. Note that result-aware metrics (e.g., FDG) measure FDC values with test results, which limits their application for test generation.

\subsection{Implementation Details}

\subsubsection{Baselines} 

We implement TfD, EntBug, DDU, and FDG for test selection using the implementation from prior work~\cite{an2022fdg}. We implement EntBug and DDU for test generation using their published EVOSUITE artifacts~\cite{ddu,entbug}. We do not include TfD for test generation since TfD is implemented on a test optimization tool utilizing bacteriologic algorithm~\cite{tfd} rather than EVOSUITE, and prior work~\cite{ddu} has shown that DDU is more powerful than TfD in terms of test generation. 

\subsubsection{Hyperparameters} To answer RQ1, RQ2, RQ3, and RQ5, we use five-fold validation\footnote{Note that some deep learning-based FL techniques~\cite{lou2021boosting,li2019deepfl,li2021fault} are evaluated in leave-one-out validation, i.e., use one buggy version as testing data and the remaining versions as training data, which is reported to cause overfitting and reduce applicability~\cite{yang2024large}. To address this concern, we use five-fold validation in the evaluation.}. In particular, we split each project's buggy versions in Defects4J (V2.0.0) into five folds: each time, one fold is selected as the testing set, and the rest four folds are used as the training set. In the training stage, we train each model for 30 epoches based on the human-written tests of the corresponding versions, resulting in our FDC metric, \techname{}. To train our model, we use a learning rate of 0.001 globally, and a batch size of 32 for memory sampling. The layer widths are (16, 16) for the embedding layer and are (16, 32, 1) for the FDC prediction network. The other parameters are empirically set as $K=10, N=100, C=20, L=5, \gamma=0.9, \sigma=0.1$. In the evaluating stage, we evaluate the performance of the four compared metrics and \techname{} on the remaining buggy versions. Following prior work~\cite{an2022fdg}, we evaluate the performance of metrics in test selection and generation. We gradually select $n$ tests for a given failing test and report the corresponding FL results in the selection scenario on human-written and automatically generated tests. We report the FL results by selecting $1, 2, \ldots, n$ tests. In the generation process, we integrate these metrics into EVOSUITE, generate tests for a given failing test, and report the corresponding FL results. In our evaluation, $n$ is set to ten by following prior work~\cite{an2022fdg} because prior work~\cite{an2022fdg} has shown that on these subjects with only ten addition tests, FL~\cite{an2022fdg} may achieve comparable performance as the full test suite (62\% of acc@1 and 80\% of acc@10).

The experiments for RQ2 and RQ3 are conducted on tests generated by EVOSUITE or Randoop, which contain randomness from the input seeds and the searching process. To ease reproducibility and control the influence of randomness, we use fixed input seeds, repeat these experiments three times, and report the average FL results. We use the \techname{} built in RQ1 for RQ2 (i.e., selecting automatically generated tests) and RQ3 (i.e., guiding EVOSUITE to generate tests).

\subsubsection{Oracle Labelling} 
The human-written tests used in RQ1, RQ4, and RQ5 have test oracles, while the EVOSUITE-generated tests used in RQ2 or RQ3 do not.
We utilize EVOSUITE in a regression way to handle the oracle problem~\cite{shamshiri2015automatically,an2022fdg}. Following prior work~\cite{an2022fdg}, we use EVOSUITE to generate tests on the buggy versions of Defects4J and execute them on the fixed versions. The tests failing on the fixed versions are labeled as ``failing tests''.

\subsubsection{EVOSUITE Configuration} We use the \texttt{generateTests} option for EVOSUITE following previous work~\cite{entbug}. For each failing test, we initially measure its coverage. Subsequently, we calculate the FDC value for each individual in the population (i.e., each generated test) using a fitness function considering their overlap with the failing test's coverage. Individuals with the highest fitness values are then selected as parents to produce offspring. In other words, for each failing test, we generate a corresponding test suite that can diagnose the fault it reveals. We set the overall search time budget as 120 seconds. Besides, we configure the \texttt{budget\_for\_each\_generation} option to 2 seconds for each fitness function evaluation, allowing the algorithm to produce 60 generations within the allocated 120 seconds.

\subsubsection{Time costs} Table~\ref{tab:time} presents the average time costs for our approach on each version of the five projects. Row ``Train'' and Row ``Test'' present our proposed approach's training and testing time. In particular, the testing time for \techname{} refers to the total time \techname{} used to predict the FDC values of all tests. From the table, the training time ranges from 3 minutes to 39 minutes, which is tolerable since training is conducted offline. The testing process costs only several seconds on average, indicating that \techname{} can efficiently predicts FDC values of tests. Overall, our approach is a lightweight RL-based technique with a small overhead.

\subsubsection{Environment} All the experiments are conducted on a workstation with 2 Intel Xeon Gold 5218R CPUs, 256GB RAM, and four 24G GPUs of GeForce RTX 3090, running Ubuntu 18.04 x64 OS. We build our experiment on PyTorch V1.8.1~\cite{PyTorch}. 



\subsection{Fault Localization Techniques} 

As the FDC metrics (including \techname{}) are proposed to boost FL, following previous work~\cite{an2022fdg}, we investigate the performance of studied FDC metrics through a widely-used Spectrum-based Fault Localization (SBFL) technique, Ochiai with aggregation~\cite{sohn2017fluccs}. SBFL is one of the most widely studied FL techniques because of its simple input and effectiveness~\cite{wong2016survey,ochiai}, and SBFL scores are often used as features for more complicated FL techniques~\cite{li2019deepfl,zou2019empirical}, which motivates us to include it as our FL technique. 

In particular, Ochiai with aggregation (abbreviated as Ochiai-agg) first computes the line-level Ochiai scores based on the coverage information of tests and then aggregates them at the method level with the highest score of all the lines in the method~\cite{sohn2017fluccs}. To measure the performance of the compared FDC metrics, we feed the original given failing test along with the tests selected or generated based on an FDC metric to Ochiai-agg, and report the FL results.

\begin{table}[tp]
\centering
\caption{Efficiency of \techname{} (seconds)}
\vspace{-0.3cm}
\begin{tabular}{l|ccccc}
\hline
Subject           & Chart     & Time     & Lang     & Math      & Closure    \\ \hline
Train          & 634.41 & 423.72 & 180.67 & 561.78 & 2,338.51 \\
Test           & 0.51   & 0.28   & 0.08    & 0.27   & 1.35     \\ \hline
\end{tabular}
\vspace{-0.3cm}
\label{tab:time}
\end{table}


\subsection{Evaluation Metrics}
Following prior work~\cite{an2022fdg,lou2021boosting,li2019deepfl,li2021fault}, we use Mean Average Precision (mAP) and Top-n Accuracy (acc@n) to evaluate the FL accuracy. 

mAP is widely used in the Information Retrieval (IR) task while also adopted for measuring FL accuracy~\cite{an2022fdg}. In our FL context, ``Average Precision'' is calculated based on the rank of multiple buggy elements, while ``Mean'' stands for averaging all the produced rankings. The calculation of mAP can be formally written as:
\begin{equation}
    mAP = \frac{1}{|F|} \sum_{f \in F} (\frac{1}{|f_B|} \sum_{b \in f_B} \frac{num\_higher\_buggy\_rank_b}{buggy\_rank_b})
\end{equation}
where $F$ denotes the set of considered buggy programs, $f_B$ denotes the set of buggy elements (i.e., buggy methods in our setting) in the buggy program $f$, $buggy\_rank_b$ denotes the rank of the buggy element $b$ in the FL produced ranking list, $num\_higher\_buggy\_rank_b$ denotes the number of buggy elements ranked higher than the buggy element $b$ (including $b$). mAP presents the overall FL accuracy considering the average performance on various buggy programs and the ranks of multiple buggy elements.

acc@n is a widely-adopted metric for measuring FL performance~\cite{lou2021boosting,li2019deepfl}. acc@n counts the number of buggy programs where at least one of the buggy elements is ranked within the top $n$ locations. The metric directly reflects how many elements developers should inspect before finding the first buggy element with FL. 

\section{Results and Analysis}
\subsection{RQ1. Human-written test selection}\label{sec:rq1}

Using human-written tests of Defects4J as the candidate test pool, we iteratively select tests with the five studied FDC metrics. In particular, for each failing test of a given buggy program (which composes an initial test suite), we iteratively select another 10 tests using an FDC metric and thus construct an augmented test suite composed of 11 human-written tests. To evaluate whether an FDC metric performs well (i.e., selecting tests to improve FL), we apply the SBFL technique Ochiai-agg to the constructed test suite and analyze its FL performance in terms of mAP and acc@n.


\begin{figure}[!htp]
    \vspace{-0.3cm}
    \centering
    \includegraphics[width=0.7\linewidth]{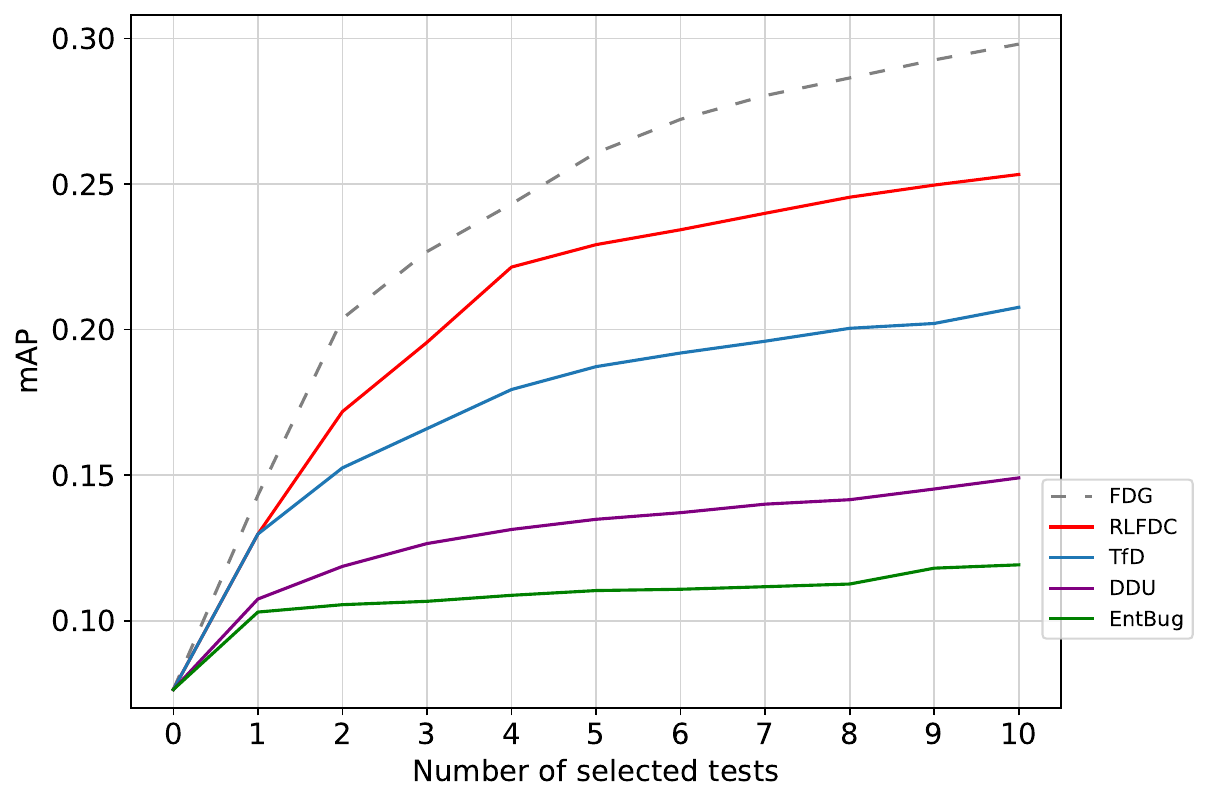}
    \vspace{-0.3cm}
    \caption{mAP values on human-written tests}
    \vspace{-0.2cm}
    \label{fig:curve}
\end{figure}

Compared with result-agnostic metrics, a result-aware metric uses extra inputs (i.e., test results), which are not always available in practice due to the widely-known test oracle problem. That is, the comparison between result-aware and result-agnostic metrics~\cite{an2022fdg} seems to be unfair. In the remaining sections, we not only conduct a fair comparison in result-agnostic metrics (i.e., comparing \techname{} against the existing result-agnostic metrics), but also conduct an unfair comparison between \techname{} and the state-of-the-art result-aware metric FDG to learn their performance gap.

Fig.~\ref{fig:curve} shows the trend of mAP values for selecting $n$ tests (where $n=1, 2, \ldots, 10$), where the horizontal axis represents the number of selected tests. As result-aware and result-agnostic metrics differ in their inputs, we distinguish them through dashed and solid lines in the figure. We observe that: (1) The FL accuracy increases with more tests. (2) With the same number of tests, the FL accuracy may vary depending on different test suites constructed by different metrics. These two observations demonstrate that the number and the FDC of tests influence FL performance to a large extent, indicating the importance of accurate FDC metrics in test selection. 

From the figure, \techname{} is the best result-agnostic metric, outperforming TfD, DDU, and EntBug. When the number of selected tests increases from 0 to 10, \techname{} consistently outperforms other result-agnostic metrics and makes the mAP value increase from 0.076 to 0.253 (an increase of 233\%). Besides, as expected, the result-aware metric FDG outperforms all the result-agnostic metrics.

\begin{table*}[!htp]
\centering
\renewcommand{\arraystretch}{0.82}
\caption{acc@n values on human-written tests}
\vspace{-0.3cm}
\resizebox{1.0\linewidth}{!}{
\begin{tabular}{cc|cccc|cccccccccccccccc}
\toprule
                           &       & @1       & @3      & @5      & @10     & @1 & @3 & @5 & \multicolumn{1}{c|}{@10} & @1 & @3 & @5  & \multicolumn{1}{c|}{@10} & @1          & @3           & @5           & \multicolumn{1}{c|}{@10}          & @1 & @3  & @5  & @10 \\ \midrule
\multicolumn{2}{c|}{Metric}        & \multicolumn{4}{c|}{EntBug(mAP=0.119)} & \multicolumn{4}{c|}{DDU(mAP=0.149)}     & \multicolumn{4}{c|}{TfD(mAP=0.208)}      & \multicolumn{4}{c|}{RLFDC(mAP=\textbf{0.253})}                                         & \multicolumn{4}{c}{\cellcolor[HTML]{E0E0E0}FDG(mAP=0.298)}    \\ \midrule
\multicolumn{1}{c|}{}      & Init. & 4        & 30      & 45      & 65      &    &    &    &                          &    &    &     &                          &             &              &              &                                   &                                                   &                                                    &                                                    & 
\\ \cmidrule{2-22} 
\multicolumn{1}{l|}{}      & 1     & 9        & 40      & 61      & 80      & 10 & 43 & 64 & \multicolumn{1}{c|}{85}  & \textbf{21} & 51 & 67  & \multicolumn{1}{c|}{84}  & \textbf{21} & \textbf{58}  & \textbf{79}  & \multicolumn{1}{c|}{\textbf{103}} & \cellcolor[HTML]{E0E0E0}{\color[HTML]{333333} 24} & \cellcolor[HTML]{E0E0E0}{\color[HTML]{333333} 64}  & \cellcolor[HTML]{E0E0E0}{\color[HTML]{333333} 87}  & \cellcolor[HTML]{E0E0E0}{\color[HTML]{333333} 109} \\
\multicolumn{1}{l|}{}      & 2     & 9        & 41      & 63      & 85      & 12 & 49 & 74 & \multicolumn{1}{c|}{99}  & 28 & 67 & 81  & \multicolumn{1}{c|}{100} & \textbf{32} & \textbf{75}  & \textbf{105} & \multicolumn{1}{c|}{\textbf{136}} & \cellcolor[HTML]{E0E0E0}{\color[HTML]{333333} 41} & \cellcolor[HTML]{E0E0E0}{\color[HTML]{333333} 100} & \cellcolor[HTML]{E0E0E0}{\color[HTML]{333333} 124} & \cellcolor[HTML]{E0E0E0}{\color[HTML]{333333} 155} \\
\multicolumn{1}{l|}{}      & 3     & 9        & 40      & 64      & 90      & 14 & 53 & 76 & \multicolumn{1}{c|}{107} & 32 & 71 & 88  & \multicolumn{1}{c|}{112} & \textbf{40} & \textbf{88}  & \textbf{112} & \multicolumn{1}{c|}{\textbf{146}} & \cellcolor[HTML]{E0E0E0}{\color[HTML]{333333} 48} & \cellcolor[HTML]{E0E0E0}{\color[HTML]{333333} 107} & \cellcolor[HTML]{E0E0E0}{\color[HTML]{333333} 135} & \cellcolor[HTML]{E0E0E0}{\color[HTML]{333333} 175} \\
\multicolumn{1}{l|}{}      & 4     & 9        & 40      & 64      & 94      & 15 & 54 & 77 & \multicolumn{1}{c|}{112} & 33 & 81 & 95  & \multicolumn{1}{c|}{127} & \textbf{46} & \textbf{104} & \textbf{129} & \multicolumn{1}{c|}{\textbf{168}} & \cellcolor[HTML]{E0E0E0}{\color[HTML]{333333} 52} & \cellcolor[HTML]{E0E0E0}{\color[HTML]{333333} 115} & \cellcolor[HTML]{E0E0E0}{\color[HTML]{333333} 145} & \cellcolor[HTML]{E0E0E0}{\color[HTML]{333333} 190} \\
\multicolumn{1}{l|}{}      & 5     & 9        & 41      & 67      & 95      & 15 & 54 & 79 & \multicolumn{1}{c|}{115} & 35 & 82 & 98  & \multicolumn{1}{c|}{130} & \textbf{47} & \textbf{108} & \textbf{137} & \multicolumn{1}{c|}{\textbf{181}} & \cellcolor[HTML]{E0E0E0}{\color[HTML]{333333} 57} & \cellcolor[HTML]{E0E0E0}{\color[HTML]{333333} 121} & \cellcolor[HTML]{E0E0E0}{\color[HTML]{333333} 158} & \cellcolor[HTML]{E0E0E0}{\color[HTML]{333333} 204} \\
\multicolumn{1}{l|}{\# Tests} & 6     & 9        & 41      & 67      & 96      & 15 & 55 & 79 & \multicolumn{1}{c|}{118} & 35 & 87 & 100 & \multicolumn{1}{c|}{134} & \textbf{47} & \textbf{112} & \textbf{141} & \multicolumn{1}{c|}{\textbf{183}} & \cellcolor[HTML]{E0E0E0}{\color[HTML]{333333} 60} & \cellcolor[HTML]{E0E0E0}{\color[HTML]{333333} 130} & \cellcolor[HTML]{E0E0E0}{\color[HTML]{333333} 164} & \cellcolor[HTML]{E0E0E0}{\color[HTML]{333333} 212} \\
\multicolumn{1}{l|}{}      & 7     & 9        & 41      & 68      & 96      & 15 & 56 & 81 & \multicolumn{1}{c|}{123} & 36 & 89 & 105 & \multicolumn{1}{c|}{142} & \textbf{48} & \textbf{115} & \textbf{144} & \multicolumn{1}{c|}{\textbf{187}} & \cellcolor[HTML]{E0E0E0}{\color[HTML]{333333} 64} & \cellcolor[HTML]{E0E0E0}{\color[HTML]{333333} 135} & \cellcolor[HTML]{E0E0E0}{\color[HTML]{333333} 170} & \cellcolor[HTML]{E0E0E0}{\color[HTML]{333333} 221} \\
\multicolumn{1}{l|}{}      & 8     & 9        & 41      & 68      & 99      & 15 & 56 & 81 & \multicolumn{1}{c|}{125} & 37 & 89 & 109 & \multicolumn{1}{c|}{145} & \textbf{49} & \textbf{117} & \textbf{149} & \multicolumn{1}{c|}{\textbf{192}} & \cellcolor[HTML]{E0E0E0}{\color[HTML]{333333} 65} & \cellcolor[HTML]{E0E0E0}{\color[HTML]{333333} 139} & \cellcolor[HTML]{E0E0E0}{\color[HTML]{333333} 174} & \cellcolor[HTML]{E0E0E0}{\color[HTML]{333333} 219} \\
\multicolumn{1}{l|}{}      & 9     & 11       & 43      & 68      & 99      & 16 & 58 & 82 & \multicolumn{1}{c|}{127} & 37 & 91 & 111 & \multicolumn{1}{c|}{149} & \textbf{49} & \textbf{121} & \textbf{155} & \multicolumn{1}{c|}{\textbf{195}} & \cellcolor[HTML]{E0E0E0}{\color[HTML]{333333} 66} & \cellcolor[HTML]{E0E0E0}{\color[HTML]{333333} 145} & \cellcolor[HTML]{E0E0E0}{\color[HTML]{333333} 178} & \cellcolor[HTML]{E0E0E0}{\color[HTML]{333333} 220} \\
\multicolumn{1}{l|}{}      & 10    & 12       & 44      & 68      & 101     & 17 & 59 & 84 & \multicolumn{1}{c|}{127} & 39 & 93 & 111 & \multicolumn{1}{c|}{149} & \textbf{50} & \textbf{124} & \textbf{156} & \multicolumn{1}{c|}{\textbf{203}} & \cellcolor[HTML]{E0E0E0}{\color[HTML]{333333} 68} & \cellcolor[HTML]{E0E0E0}{\color[HTML]{333333} 148} & \cellcolor[HTML]{E0E0E0}{\color[HTML]{333333} 179} & \cellcolor[HTML]{E0E0E0}{\color[HTML]{333333} 222} \\ \cmidrule{2-22} 
\multicolumn{1}{l|}{}      & Full  & 110      & 208     & 250     & 277     &    &    &    &                          &    &    &     &                          &             &              &              &                                   &                                                   &                                                    &                                                    &                                                    \\ \bottomrule
\end{tabular}
}
\vspace{-0.2cm}
\label{tab:rq1}
\end{table*}

We further present the acc@n values of Ochiai-agg based on human-written tests selected by each metric in Table~\ref{tab:rq1}. The last row presents the FL results using all human-written tests of each subject. As the proposed \techname{} is a result-agnostic metric, for any number of selected tests, we highlight the best results among the four result-agnostic metrics (i.e., EntBug, DDU, TfD, and \techname{}) with the bold font. The last four columns give the acc@n values of the result-aware metric FDG for reference. The mAP values for selecting ten tests are also shown in the brackets next to the metric names. From the table, \techname{} performs the best among all result-agnostic metrics, achieving the highest acc@n (n=1, 3, 5, 10) when selecting any number of tests and the highest mAP. In particular, \techname{} achieves 50 in terms of acc@1 and 203 in terms of acc@10 when selecting ten tests. Compared to the second-best result-agnostic metric, TfD, the improvement on acc@1 and acc@10 is 28.2\% and 36.2\%, respectively. With ten tests selected by \techname{}, the acc@1, acc@3, acc@5, and acc@10 improve 11.5 times, 3.13 times, 2.47 times, and 2.12 times respectively compared to the initial test suite. When compared to the full test suite (which contains over 1K tests), the ten tests selected by \techname{}~achieve 45.4\% and 73.3\% of its acc@1 and acc@10 values, respectively.

\begin{figure}
    \centering
    \includegraphics[width=0.7\textwidth]{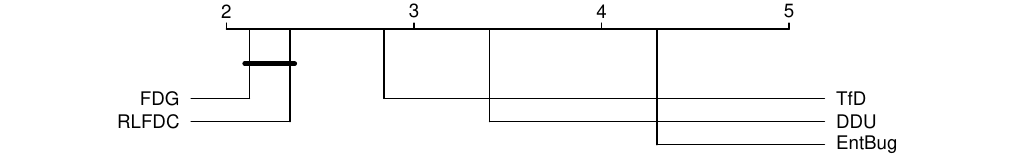}
    \caption{Critical difference plot}
    \vspace{-0.3cm}
    \label{fig:statistical_test}
\end{figure}

To investigate whether these metrics perform significantly differently in terms of mAP, we conduct a non-parametric ANOVA analysis utilizing the Friedman test with the Iman and Davenport extension~\cite{iman1980approximations}, which is robust to non-normality. The analysis result rejects the null hypothesis of equal performance (with $p<2.2\times10^{-16}$), indicating that at least one metric performs significantly differently from the others. Furthermore, to investigate which metrics perform significantly better, we run a post hoc test (i.e., Friedman post hoc test with Bergmann and Hommel's correction~\cite{calvo2016scmamp}) to compare each pair of metrics. The analysis results are given by Fig.~\ref{fig:statistical_test}, where each metric is placed on an axis according to its mean rank among the five metrics across all outcomes. Metrics with larger mAP are placed on the left. The metrics not grouped with a horizontal line are significantly different ($p < 0.05$). From the figure, the proposed result-agnostic metric \techname{} and the result-aware metric FDG do not perform significantly differently, but each of them significantly outperforms other metrics.

\begin{table}[htp]
\renewcommand{\arraystretch}{0.82}
\caption{acc@n values on human-written tests of ten additional subjects}
\vspace{-0.2cm}
\resizebox{1.0\linewidth}{!}{
\begin{tabular}{cc|cccc|cccccccccccccccc}
\toprule
                              &       & @1      & @3       & @5      & @10     & @1 & @3 & @5  & \multicolumn{1}{c|}{@10} & @1          & @3 & @5  & \multicolumn{1}{c|}{@10} & @1          & @3           & @5           & \multicolumn{1}{c|}{@10}          & @1                         & @3                          & @5                          & @10                         \\ \midrule
\multicolumn{2}{c|}{Metric}           & \multicolumn{4}{c|}{EntBug(mAP=0.175)} & \multicolumn{4}{c|}{DDU(mAP=0.246)}      & \multicolumn{4}{c|}{TfD(mAP=0.237)}               & \multicolumn{4}{c|}{RLFDC(mAP=\textbf{0.285})}                                         & \multicolumn{4}{c}{\cellcolor[HTML]{E0E0E0}FDG(mAP=0.312)}                                                           \\ \midrule
\multicolumn{1}{c|}{}         & Init. & 4       & 14       & 18      & 29      &    &    &     &                          &             &    &     &                          &             &              &              &                                   &                            &                             &                             &                             \\ \cmidrule{2-22} 
\multicolumn{1}{c|}{}         & 1     & 8       & 23       & 37      & 64      & 8  & 18 & 25  & \multicolumn{1}{c|}{46}  & \textbf{15} & 30 & 41  & \multicolumn{1}{c|}{58}  & 14          & \textbf{34}  & \textbf{48}  & \multicolumn{1}{c|}{\textbf{80}}  & \cellcolor[HTML]{E0E0E0}12 & \cellcolor[HTML]{E0E0E0}31  & \cellcolor[HTML]{E0E0E0}40  & \cellcolor[HTML]{E0E0E0}67  \\
\multicolumn{1}{c|}{}         & 2     & 10      & 27       & 42      & 71      & 19 & 32 & 53  & \multicolumn{1}{c|}{86}  & 21          & 35 & 52  & \multicolumn{1}{c|}{85}  & \textbf{24} & \textbf{52}  & \textbf{75}  & \multicolumn{1}{c|}{\textbf{105}} & \cellcolor[HTML]{E0E0E0}18 & \cellcolor[HTML]{E0E0E0}49  & \cellcolor[HTML]{E0E0E0}67  & \cellcolor[HTML]{E0E0E0}94  \\
\multicolumn{1}{c|}{}         & 3     & 13      & 30       & 50      & 80      & 26 & 49 & 64  & \multicolumn{1}{c|}{99}  & 26          & 46 & 61  & \multicolumn{1}{c|}{94}  & \textbf{30} & \textbf{64}  & \textbf{91}  & \multicolumn{1}{c|}{\textbf{119}} & \cellcolor[HTML]{E0E0E0}26 & \cellcolor[HTML]{E0E0E0}64  & \cellcolor[HTML]{E0E0E0}85  & \cellcolor[HTML]{E0E0E0}119 \\
\multicolumn{1}{c|}{}         & 4     & 17      & 36       & 59      & 90      & 30 & 55 & 77  & \multicolumn{1}{c|}{108} & 27          & 52 & 70  & \multicolumn{1}{c|}{104} & \textbf{35} & \textbf{72}  & \textbf{96}  & \multicolumn{1}{c|}{\textbf{127}} & \cellcolor[HTML]{E0E0E0}40 & \cellcolor[HTML]{E0E0E0}80  & \cellcolor[HTML]{E0E0E0}103 & \cellcolor[HTML]{E0E0E0}146 \\
\multicolumn{1}{c|}{}         & 5     & 18      & 37       & 60      & 93      & 32 & 63 & 84  & \multicolumn{1}{c|}{119} & 31          & 60 & 83  & \multicolumn{1}{c|}{110} & \textbf{37} & \textbf{74}  & \textbf{95}  & \multicolumn{1}{c|}{\textbf{133}} & \cellcolor[HTML]{E0E0E0}43 & \cellcolor[HTML]{E0E0E0}94  & \cellcolor[HTML]{E0E0E0}115 & \cellcolor[HTML]{E0E0E0}156 \\
\multicolumn{1}{c|}{\# Tests} & 6     & 18      & 39       & 62      & 96      & 34 & 64 & 87  & \multicolumn{1}{c|}{124} & 36          & 66 & 87  & \multicolumn{1}{c|}{120} & \textbf{39} & \textbf{80}  & \textbf{104} & \multicolumn{1}{c|}{\textbf{140}} & \cellcolor[HTML]{E0E0E0}50 & \cellcolor[HTML]{E0E0E0}100 & \cellcolor[HTML]{E0E0E0}121 & \cellcolor[HTML]{E0E0E0}160 \\
\multicolumn{1}{c|}{}         & 7     & 18      & 41       & 64      & 99      & 35 & 68 & 92  & \multicolumn{1}{c|}{128} & 36          & 68 & 92  & \multicolumn{1}{c|}{126} & \textbf{41} & \textbf{87}  & \textbf{108} & \multicolumn{1}{c|}{\textbf{144}} & \cellcolor[HTML]{E0E0E0}49 & \cellcolor[HTML]{E0E0E0}106 & \cellcolor[HTML]{E0E0E0}129 & \cellcolor[HTML]{E0E0E0}164 \\
\multicolumn{1}{c|}{}         & 8     & 20      & 48       & 69      & 106     & 36 & 70 & 95  & \multicolumn{1}{c|}{134} & 36          & 66 & 92  & \multicolumn{1}{c|}{130} & \textbf{46} & \textbf{90}  & \textbf{113} & \multicolumn{1}{c|}{\textbf{151}} & \cellcolor[HTML]{E0E0E0}50 & \cellcolor[HTML]{E0E0E0}108 & \cellcolor[HTML]{E0E0E0}133 & \cellcolor[HTML]{E0E0E0}167 \\
\multicolumn{1}{c|}{}         & 9     & 20      & 48       & 70      & 111     & 39 & 74 & 101 & \multicolumn{1}{c|}{137} & 36          & 68 & 98  & \multicolumn{1}{c|}{137} & \textbf{47} & \textbf{95}  & \textbf{114} & \multicolumn{1}{c|}{\textbf{152}} & \cellcolor[HTML]{E0E0E0}50 & \cellcolor[HTML]{E0E0E0}112 & \cellcolor[HTML]{E0E0E0}136 & \cellcolor[HTML]{E0E0E0}172 \\
\multicolumn{1}{c|}{}         & 10    & 22      & 50       & 72      & 117     & 40 & 76 & 103 & \multicolumn{1}{c|}{139} & 38          & 74 & 100 & \multicolumn{1}{c|}{140} & \textbf{46} & \textbf{100} & \textbf{119} & \multicolumn{1}{c|}{\textbf{156}} & \cellcolor[HTML]{E0E0E0}54 & \cellcolor[HTML]{E0E0E0}112 & \cellcolor[HTML]{E0E0E0}138 & \cellcolor[HTML]{E0E0E0}173 \\ \cmidrule{2-22} 
\multicolumn{1}{c|}{}         & Full  & 80      & 143      & 167     & 196     &    &    &     &                          &             &    &     &                          &             &              &              &                                   &                            &                             &                             &                             \\ \bottomrule
\end{tabular}
\label{tab:add_sub}
}
\end{table}

To further validate the generalizability of RLFDC, apart from the five subjects widely used for evaluating fault localization techniques~\cite{kang2024quantitative,li2019deepfl,zou2019empirical,xie2022universal,li2021fault}, we also include ten additional subjects from Defects4J V2.0.0 in this RQ for further evaluation. We conduct an experiment with ten additional subjects, encompassing a total of 264 bugs, from Defects4J V2.0.0\footnote{We exclude the first 45 bugs in Jsoup and all bugs in Gson/JacksonCore because of reproducing problem. This problem can also be seen in previous work~\cite{lou2021boosting,yang2024large}.}. Table~\ref{tab:add_sub} presents how RLFDC and the four baselines perform in human-written test selection for the ten subjects. 
From the table, RLFDC consistently outperforms all other result-agnostic metrics, achieving the highest acc@n (n=1, 3, 5, 10) across various number of selected tests, as well as the highest mAP.
In particular, \techname{} achieves 46 in terms of acc@1 and 156 in terms of acc@10 when selecting ten tests. Compared to the second-best result-agnostic metric, TfD, the improvement on acc@1 and acc@10 is 21.1\% and 11.4\%, respectively. With ten tests selected by \techname{}, the acc@1, acc@3, acc@5, and acc@10 improve 10.5 times, 6.1 times, 5.6 times, and 4.4 times respectively compared to the initial test suite. When compared to the full test suite, the ten tests selected by \techname{}~achieve 57.5\% and 79.6\% of its acc@1 and acc@10 values, respectively.

\subsection{RQ2. Automatically generated test selection}
\label{sec:rq2}

\begin{table*}[!htp]
\centering
\renewcommand{\arraystretch}{0.82}
\caption{acc@n values on tests automatically generated by EVOSUITE}
\vspace{-0.3cm}
\resizebox{1.0\linewidth}{!}{
\begin{tabular}{cc|cccc|cccccccccccccccc}
\toprule
                           &       & @1       & @3      & @5      & @10     & @1 & @3 & @5 & \multicolumn{1}{c|}{@10} & @1 & @3 & @5  & \multicolumn{1}{c|}{@10} & @1          & @3           & @5           & \multicolumn{1}{c|}{@10}          & @1 & @3  & @5  & @10 \\ \midrule
\multicolumn{2}{c|}{Metric}        & \multicolumn{4}{c|}{EntBug(mAP=0.215)}  & \multicolumn{4}{c|}{DDU(mAP=0.230)}                                        & \multicolumn{4}{c|}{TfD(mAP=0.226)}     & \multicolumn{4}{c|}{RLFDC(mAP=\textbf{0.236})}                                       & \multicolumn{4}{c}{\cellcolor[HTML]{E0E0E0}FDG(mAP=0.239)}                                                         \\ \midrule
\multicolumn{1}{c|}{}      & Init. & 4  & 26 & 39 & \multicolumn{1}{c|}{57}   &    &    &    &                          &    &    &     &                          &             &              &              &                                   &                                                   &                                                    &                                                    & 
\\ \cmidrule{2-22} 
\multicolumn{1}{c|}{}      & 1     & 16 & 47 & 58 & \multicolumn{1}{c|}{75}  & \textbf{19} & \textbf{51} & \textbf{62} & \multicolumn{1}{c|}{\textbf{77}} & 16 & 40 & 53 & \multicolumn{1}{c|}{73}  & 14          & 49          & 60          & \multicolumn{1}{c|}{76}           & \cellcolor[HTML]{E0E0E0}20 & \cellcolor[HTML]{E0E0E0}52 & \cellcolor[HTML]{E0E0E0}66 & \cellcolor[HTML]{E0E0E0}84  \\
\multicolumn{1}{c|}{}      & 2     & 20 & 53 & 63 & \multicolumn{1}{c|}{80}  & \textbf{26} & \textbf{56} & \textbf{70} & \multicolumn{1}{c|}{86}          & 22 & 46 & 62 & \multicolumn{1}{c|}{78}  & 23          & 55          & 68          & \multicolumn{1}{c|}{\textbf{88}}  & \cellcolor[HTML]{E0E0E0}26 & \cellcolor[HTML]{E0E0E0}59 & \cellcolor[HTML]{E0E0E0}74 & \cellcolor[HTML]{E0E0E0}95  \\
\multicolumn{1}{c|}{}      & 3     & 24 & 53 & 64 & \multicolumn{1}{c|}{86}  & \textbf{29} & \textbf{58} & \textbf{72} & \multicolumn{1}{c|}{91}          & 25 & 50 & 63 & \multicolumn{1}{c|}{84}  & 27          & 57          & 71          & \multicolumn{1}{c|}{\textbf{97}}  & \cellcolor[HTML]{E0E0E0}31 & \cellcolor[HTML]{E0E0E0}63 & \cellcolor[HTML]{E0E0E0}78 & \cellcolor[HTML]{E0E0E0}104 \\
\multicolumn{1}{c|}{}      & 4     & 25 & 55 & 64 & \multicolumn{1}{c|}{89}  & \textbf{30} & 59          & 73          & \multicolumn{1}{c|}{96}          & 26 & 55 & 67 & \multicolumn{1}{c|}{87}  & 29          & \textbf{60} & \textbf{74} & \multicolumn{1}{c|}{\textbf{101}} & \cellcolor[HTML]{E0E0E0}34 & \cellcolor[HTML]{E0E0E0}65 & \cellcolor[HTML]{E0E0E0}81 & \cellcolor[HTML]{E0E0E0}110 \\
\multicolumn{1}{c|}{}      & 5     & 26 & 56 & 65 & \multicolumn{1}{c|}{91}  & 30          & 60          & 73          & \multicolumn{1}{c|}{98}          & 29 & 57 & 69 & \multicolumn{1}{c|}{89}  & \textbf{31} & \textbf{62} & \textbf{76} & \multicolumn{1}{c|}{\textbf{104}} & \cellcolor[HTML]{E0E0E0}34 & \cellcolor[HTML]{E0E0E0}67 & \cellcolor[HTML]{E0E0E0}83 & \cellcolor[HTML]{E0E0E0}112 \\
\multicolumn{1}{c|}{\# Tests} & 6     & 26 & 57 & 67 & \multicolumn{1}{c|}{92}  & 31          & 62          & 75          & \multicolumn{1}{c|}{100}         & 30 & 59 & 70 & \multicolumn{1}{c|}{93}  & \textbf{33} & \textbf{63} & \textbf{78} & \multicolumn{1}{c|}{\textbf{105}} & \cellcolor[HTML]{E0E0E0}34 & \cellcolor[HTML]{E0E0E0}67 & \cellcolor[HTML]{E0E0E0}83 & \cellcolor[HTML]{E0E0E0}114 \\
\multicolumn{1}{c|}{}      & 7     & 27 & 58 & 68 & \multicolumn{1}{c|}{94}  & 31          & 63          & 76          & \multicolumn{1}{c|}{103}         & 31 & 60 & 71 & \multicolumn{1}{c|}{96}  & \textbf{33} & \textbf{65} & \textbf{80} & \multicolumn{1}{c|}{\textbf{108}} & \cellcolor[HTML]{E0E0E0}34 & \cellcolor[HTML]{E0E0E0}67 & \cellcolor[HTML]{E0E0E0}84 & \cellcolor[HTML]{E0E0E0}114 \\
\multicolumn{1}{c|}{}      & 8     & 28 & 58 & 69 & \multicolumn{1}{c|}{95}  & 32          & 63          & 78          & \multicolumn{1}{c|}{106}         & 32 & 62 & 73 & \multicolumn{1}{c|}{101} & \textbf{33} & \textbf{66} & \textbf{81} & \multicolumn{1}{c|}{\textbf{111}} & \cellcolor[HTML]{E0E0E0}34 & \cellcolor[HTML]{E0E0E0}67 & \cellcolor[HTML]{E0E0E0}84 & \cellcolor[HTML]{E0E0E0}113 \\
\multicolumn{1}{c|}{}      & 9     & 29 & 59 & 71 & \multicolumn{1}{c|}{98}  & 33          & 64          & 79          & \multicolumn{1}{c|}{107}         & 33 & 63 & 76 & \multicolumn{1}{c|}{105} & \textbf{34} & \textbf{67} & \textbf{83} & \multicolumn{1}{c|}{\textbf{113}} & \cellcolor[HTML]{E0E0E0}35 & \cellcolor[HTML]{E0E0E0}67 & \cellcolor[HTML]{E0E0E0}83 & \cellcolor[HTML]{E0E0E0}113 \\
\multicolumn{1}{l|}{}      & 10    & 30 & 59 & 72 & \multicolumn{1}{c|}{100} & 33          & 64          & 80          & \multicolumn{1}{c|}{107}         & 33 & 63 & 76 & \multicolumn{1}{c|}{107} & \textbf{34} & \textbf{68} & \textbf{84} & \multicolumn{1}{c|}{\textbf{113}} & \cellcolor[HTML]{E0E0E0}35 & \cellcolor[HTML]{E0E0E0}68 & \cellcolor[HTML]{E0E0E0}84 & \cellcolor[HTML]{E0E0E0}113 \\ \cmidrule{2-22} 
\multicolumn{1}{l|}{}      & Full  & 37 & 70 & 88 & \multicolumn{1}{c|}{117} &             &             &             &                                  &    &    &    &                          &             &             &             &                                   &                            &                            &                            &                             \\ \bottomrule
\end{tabular}
}
\vspace{-0.2cm}
\label{tab:rq2}
\end{table*}


In this section, we still evaluate the studied metrics in the test selection task, but the candidate test set is automatically generated by a test generation tool. 
This practical scenario aims to balance the effort of human labelling and FL performance, where FDC metrics are utilized to select tests from the generated test suite.
In particular, for each buggy program, we use EVOSUITE or Randoop to generate tests for 120 seconds. The generated tests are then used as a candidate test pool. For each failing test, we iteratively select 10 tests from the generated test suite based on a studied metric, then use the 11 tests as the input of Ochiai-agg.

Table~\ref{tab:rq2} presents the acc@n values with selected EVOSUITE-generated tests. From the table, on EVOSUITE-generated tests, \techname{} outperforms all result-agnostic metrics in terms of acc@n (n=1, 3, 5, 10) and mAP when selecting more than three tests. With ten tests selected by \techname{}, the acc@1 and acc@10 improve 7.5 times and 1.0 times respectively compared to the initial test suite. Moreover, the ten tests selected by \techname{} achieve 91.9\% and 96.6\% of the full test suite's acc@1 and acc@10, respectively, indicating that \techname{} helps developers save the cost of labeling oracles while still maintaining satisfactory FL performance by selecting tests with high FDC values.
It is also worth noting that on EVOSUITE-generated tests, \techname{} has competitive performance with the result-aware metric FDG. Compared with Table~\ref{tab:rq1}, the performance of different metrics decreases and the advantage of \techname{} over other metrics becomes less obvious. This may be attributed to the quality of automatically generated tests. 
The quality of automatically generated tests often falls short of that produced by human-written tests. Developers possess a deep understanding of their projects, including knowledge of error-prone inputs and suspicious conditions, a grasp of each method's functionality, and an awareness of the correct sequence for invoking methods. In contrast, test generation tools lack this intimate project understanding. Consequently, they might generate irrelevant inputs, ineffective invocations, and call methods in an incorrect sequence, which can impair the fault diagnosis capability of the generated tests.
Therefore, there is a lack of tests with high FDC values available for selection, which is also reflected by the steep fall of FL performance using the full test suite.

\begin{table}[htp]
\renewcommand{\arraystretch}{0.82}
\caption{acc@n values on tests automatically generated by Randoop}
\vspace{-0.3cm}
\resizebox{1.0\linewidth}{!}{
\begin{tabular}{cc|cccc|cccccccccccccccc}
\toprule
                              &       & @1      & @3      & @5      & @10      & @1 & @3          & @5          & \multicolumn{1}{c|}{@10}         & @1          & @3          & @5 & \multicolumn{1}{c|}{@10} & @1          & @3          & @5          & \multicolumn{1}{c|}{@10}         & @1                         & @3                         & @5                         & @10                        \\ \midrule
\multicolumn{2}{c|}{Metric}           & \multicolumn{4}{c|}{EntBug(mAP=0.239)} & \multicolumn{4}{c|}{DDU(mAP=0.291)}                               & \multicolumn{4}{c|}{TfD(mAP=0.295)}                       & \multicolumn{4}{c|}{RLFDC(mAP=\textbf{0.305})}                                      & \multicolumn{4}{c}{\cellcolor[HTML]{E0E0E0}FDG(mAP=0.302)}                                                        \\ \midrule
\multicolumn{1}{c|}{}         & Init. & 2       & 22      & 28      & 43       &    &             &             &                                  &             &             &    &                          &             &             &             &                                  &                            &                            &                            &                            \\ \cmidrule{2-22} 
\multicolumn{1}{c|}{}         & 1     & 10      & 32      & 41      & 52       & 13 & \textbf{33} & \textbf{44} & \multicolumn{1}{c|}{\textbf{55}} & \textbf{14} & \textbf{33} & 39 & \multicolumn{1}{c|}{51}  & 11          & \textbf{33} & 43          & \multicolumn{1}{c|}{52}          & \cellcolor[HTML]{E0E0E0}17 & \cellcolor[HTML]{E0E0E0}36 & \cellcolor[HTML]{E0E0E0}46 & \cellcolor[HTML]{E0E0E0}54 \\
\multicolumn{1}{c|}{}         & 2     & 11      & 33      & 44      & 53       & 18 & \textbf{39} & \textbf{50} & \multicolumn{1}{c|}{\textbf{58}} & \textbf{19} & 38          & 45 & \multicolumn{1}{c|}{55}  & \textbf{19} & 38          & 48          & \multicolumn{1}{c|}{\textbf{58}} & \cellcolor[HTML]{E0E0E0}23 & \cellcolor[HTML]{E0E0E0}42 & \cellcolor[HTML]{E0E0E0}52 & \cellcolor[HTML]{E0E0E0}59 \\
\multicolumn{1}{c|}{}         & 3     & 11      & 33      & 44      & 54       & 19 & \textbf{42} & \textbf{52} & \multicolumn{1}{c|}{\textbf{60}} & 21          & 39          & 48 & \multicolumn{1}{c|}{56}  & \textbf{22} & \textbf{42} & 50          & \multicolumn{1}{c|}{\textbf{60}} & \cellcolor[HTML]{E0E0E0}24 & \cellcolor[HTML]{E0E0E0}45 & \cellcolor[HTML]{E0E0E0}53 & \cellcolor[HTML]{E0E0E0}61 \\
\multicolumn{1}{c|}{}         & 4     & 12      & 34      & 44      & 54       & 20 & \textbf{43} & \textbf{54} & \multicolumn{1}{c|}{61}          & 22          & 41          & 51 & \multicolumn{1}{c|}{58}  & \textbf{25} & \textbf{43} & 53          & \multicolumn{1}{c|}{\textbf{62}} & \cellcolor[HTML]{E0E0E0}24 & \cellcolor[HTML]{E0E0E0}47 & \cellcolor[HTML]{E0E0E0}54 & \cellcolor[HTML]{E0E0E0}63 \\
\multicolumn{1}{c|}{}         & 5     & 13      & 34      & 44      & 55       & 20 & \textbf{44} & 53          & \multicolumn{1}{c|}{61}          & 23          & 42          & 52 & \multicolumn{1}{c|}{59}  & \textbf{25} & \textbf{44} & \textbf{55} & \multicolumn{1}{c|}{\textbf{64}} & \cellcolor[HTML]{E0E0E0}24 & \cellcolor[HTML]{E0E0E0}46 & \cellcolor[HTML]{E0E0E0}55 & \cellcolor[HTML]{E0E0E0}64 \\
\multicolumn{1}{c|}{\# Tests} & 6     & 14      & 34      & 45      & 55       & 21 & \textbf{45} & 54          & \multicolumn{1}{c|}{62}          & 23          & 43          & 53 & \multicolumn{1}{c|}{60}  & \textbf{24} & \textbf{45} & \textbf{56} & \multicolumn{1}{c|}{\textbf{66}} & \cellcolor[HTML]{E0E0E0}23 & \cellcolor[HTML]{E0E0E0}47 & \cellcolor[HTML]{E0E0E0}56 & \cellcolor[HTML]{E0E0E0}66 \\
\multicolumn{1}{c|}{}         & 7     & 14      & 35      & 45      & 55       & 23 & \textbf{45} & 54          & \multicolumn{1}{c|}{63}          & 23          & 44          & 53 & \multicolumn{1}{c|}{60}  & \textbf{25} & \textbf{45} & \textbf{56} & \multicolumn{1}{c|}{\textbf{65}} & \cellcolor[HTML]{E0E0E0}24 & \cellcolor[HTML]{E0E0E0}48 & \cellcolor[HTML]{E0E0E0}56 & \cellcolor[HTML]{E0E0E0}66 \\
\multicolumn{1}{c|}{}         & 8     & 16      & 35      & 45      & 56       & 23 & 45          & 54          & \multicolumn{1}{c|}{63}          & 24          & 45          & 53 & \multicolumn{1}{c|}{61}  & \textbf{26} & \textbf{46} & \textbf{56} & \multicolumn{1}{c|}{\textbf{66}} & \cellcolor[HTML]{E0E0E0}26 & \cellcolor[HTML]{E0E0E0}48 & \cellcolor[HTML]{E0E0E0}56 & \cellcolor[HTML]{E0E0E0}67 \\
\multicolumn{1}{c|}{}         & 9     & 16      & 36      & 45      & 56       & 23 & \textbf{46} & 54          & \multicolumn{1}{c|}{64}          & 25          & \textbf{46} & 53 & \multicolumn{1}{c|}{62}  & \textbf{27} & \textbf{46} & \textbf{56} & \multicolumn{1}{c|}{\textbf{67}} & \cellcolor[HTML]{E0E0E0}26 & \cellcolor[HTML]{E0E0E0}48 & \cellcolor[HTML]{E0E0E0}55 & \cellcolor[HTML]{E0E0E0}67 \\
\multicolumn{1}{c|}{}         & 10    & 17      & 36      & 45      & 56       & 24 & 46          & 55          & \multicolumn{1}{c|}{64}          & 25          & 46          & 54 & \multicolumn{1}{c|}{62}  & \textbf{28} & \textbf{47} & \textbf{56} & \multicolumn{1}{c|}{\textbf{67}} & \cellcolor[HTML]{E0E0E0}26 & \cellcolor[HTML]{E0E0E0}49 & \cellcolor[HTML]{E0E0E0}56 & \cellcolor[HTML]{E0E0E0}67 \\ \cmidrule{2-22} 
\multicolumn{1}{c|}{}         & Full  & 30      & 48      & 59      & 71       &    &             &             &                                  &             &             &    &                          &             &             &             &                                  &                            &                            &                            &                            \\ \bottomrule
\end{tabular}
\label{tab:rq2-randoop}
}
\end{table}

Since EVOSUITE cannot generate compilable tests for 103 faults, to mitigate this limitation and investigate the generalizability of FDC metrics, we employed Randoop~\cite{pacheco2007randoop} across the five projects to generate tests. Specifically, we utilized the \texttt{gen\_tests.pl} script provided by Defects4J to generate tests for relevant classes with a total time budget of 120 seconds. We conducted the experiment three times using different random seeds and reported the averaged results. Overall, Randoop generates compilable tests for 133 faults, and the FL results using these tests are displayed in Table~\ref{tab:rq2-randoop}. From the table, RLFDC continues to outperform all other result-agnostic metrics in terms of acc@n and mAP. When compared to the result-aware metric FDG, RLFDC not only achieves competitive acc@n but also a higher mAP (0.305 vs 0.302). Notably, with ten tests selected by \techname{}, the acc@1 and acc@10 improve 13.0 times and 0.6 times respectively, compared to the initial test suite. Furthermore, the ten tests selected by \techname{} achieve 93.3\% and 94.4\% of the full test suite's acc@1 and acc@10, respectively.

\subsection{RQ3. Automated test generation}

In this section, we evaluate the studied metrics by using them as guidance for test generation. EVOSUITE is a search-based test generation tool based on an evolutionary algorithm, which uses a fitness function to guide the evolution direction. EVOSUITE has several default fitness functions (e.g., line coverage). As this paper targets FL, we replace the fitness function of EVOSUITE with the studied metrics (i.e., DDU, EntBug, and \techname{}), and investigate whether the resulting test generation techniques (abbreviated as EVO-DDU, EVO-EntBug, EVO-\techname{}) can generate tests with high FDC. Besides, we include another baseline, i.e., the EVOSUITE with line coverage as the fitness function, which we refer to as EVO-line. Note that in RQ3 we discard TfD and FDG because prior work~\cite{ddu} shows that DDU is more powerful than TfD in terms of test generation. The result-aware metric FDG cannot be applied in this study because FDG requires test oracles, which is impractical to get for the whole population during the evolution process. 

We compare the FL performance of all tests generated based on various FDC metrics. In particular, we run each generation technique, i.e., EVO-line, EVO-DDU, EVO-EntBug, and EVO-\techname{}, for 120 seconds and feed the corresponding generated test suite to Ochiai-agg. Their FL results are given by Row ``Full'' in Table~\ref{tab:rq3}. According to this row, EVO-\techname{} achieves an improvement of 48.6\% and 8.5\% in terms of acc@1 and acc@10 compared to EVO-line, and an improvement of 3.8\% and 1.6\% in terms of acc@1 and acc@10 compared to the state-of-the-art technique, DDU.  That is, \techname{} outperforms all result-agnostic metrics in terms of acc@n (n=1, 3, 5, 10) when generating tests for FL within a given time limit. However, we also observe that these generation techniques generate various numbers of tests within 120 seconds, i.e., EVO-line, EVO-DDU, EVO-EntBug, and EVO-\techname{} generate 42,482, 24,347, 13,026, and 24,271 tests respectively. We wonder whether the various number of generated tests would influence the FL performance. 

To address this concern, we conduct another experiment to align the number of tests used for FL. In particular, for each of these generated test suites, we iteratively select ten tests with the highest FDC values\footnote{Here we use \techname{} as the selection criterion because \techname{} is demonstrated to be the best result-agnostic metric according to RQ1 and RQ2.}, and then feed the selected ten tests and the initial failing test to Ochiai-agg. The corresponding results are given by Row ``10'' in Table~\ref{tab:rq3}.
From the table, \techname{} again outperforms all result-agnostic metrics in terms of both acc@n (n=1, 3, 5, 10) and mAP (given in brackets) in test generation for FL, even if we control the number of tests. In particular, with ten tests, EVO-\techname{} successfully helps the FL technique to localize 9 more faults within top-1, and 8 more faults within top-10 compared to EVO-line, and achieves an improvement of 4.9\% and 3.1\% in terms of acc@1 and mAP compared to EVO-DDU.
In summary, the experimental results demonstrate that \techname{} can help guide the EVOSUITE to generate tests with high FDC and enhance the upper bound of FL.

\begin{table*}[tp]
\centering
\renewcommand{\arraystretch}{0.60}
\caption{acc@n values on tests generated under the guidance of different metrics}
\vspace{-0.3cm}
\resizebox{1.0\linewidth}{!}{
\begin{tabular}{cc|cccccccccccccccc}
\toprule

                           &       & @1  & @3 & @5 & \multicolumn{1}{c|}{@10} & @1 & @3 & @5  & \multicolumn{1}{c|}{@10}          & @1 & @3 & @5 & \multicolumn{1}{c|}{@10} & @1          & @3          & @5           & @10          \\ \midrule
\multicolumn{2}{c|}{Metric}        & \multicolumn{4}{c|}{EVO-line(mAP=0.236)} & \multicolumn{4}{c|}{EVO-DDU(mAP=0.258)}               & \multicolumn{4}{c|}{EVO-EntBug(mAP=0.261)}  & \multicolumn{4}{c}{EVO-\techname{}(mAP=\textbf{0.266})}                    \\ \midrule
\multicolumn{1}{c|}{}      & Init. & 4   & 26 & 39 & \multicolumn{1}{c|}{57}  & & & & & & & & & & & &
\\ \cmidrule{2-18} 
\multicolumn{1}{c|}{\# Tests} & 10    & 34  & 68 & 84 & \multicolumn{1}{c|}{113} & 41 & 75 & 96  & \multicolumn{1}{c|}{\textbf{121}} & 41 & 76 & 94 & \multicolumn{1}{c|}{117} & \textbf{43} & \textbf{81} & \textbf{98}  & \textbf{121} \\ \cmidrule{2-18} 
\multicolumn{1}{c|}{}      & Full  & 37  & 70 & 88 & \multicolumn{1}{c|}{117} & 53 & 85 & 102 & \multicolumn{1}{c|}{125}          & 50 & 81 & 99 & \multicolumn{1}{c|}{122} & \textbf{55} & \textbf{86} & \textbf{103} & \textbf{127} \\ \bottomrule
\end{tabular}
}
\label{tab:rq3}
\end{table*}


\begin{table}[tp]
\centering
\caption{acc@10 values in cross-project scenario}
\vspace{-0.3cm}
\begin{tabular}{l|l|ccccc}
\hline
Train Subj. & Metrics             & Chart & Time & Lang & Math & Closure \\ \hline
Chart            & \multirow{5}{*}{\techname{}} & -     & 12   & 58   & 80   & 28      \\
Time             &                        & 20    & -    & 55   & 79   & 26      \\
Lang             &                        & 20    & 8    & -    & 83   & 24      \\
Math             &                        & 18    & 9    & 58   & -    & 25      \\
Closure          &                        & 19    & 14   & 54   & 76   & -       \\ \hline
-                & \techname{}            & 21    & 15   & 57   & 81   & 29      \\
-                & TfD                    & 17    & 7    & 53   & 52   & 21      \\  \hline
\end{tabular}
\vspace{-0.2cm}
\label{tab:rq4}
\end{table}

\subsection{RQ4. In the cross-project scenario}
Our approach builds a better metric \techname{} through a training process, which can not only be conducted in cross-version scenario (RQ1-3), but also be conducted in cross-project scenario. To evaluate \techname{} in the latter scenario, for each project, we train a RL model using the human-written tests, resulting in five different \techname{}. Each \techname{} is applied to human-written test selection for FL on other four projects, and the corresponding FL results are reported by Table~\ref{tab:rq4}. Due to space limitation, we only report the acc@10 values. The first column presents the five subjects used for training. The 1st to 5th rows present the performance of \techname{} trained on these five subjects respectively. The 6th row presents the performance of \techname{} when being trained and used in different versions of a project as in RQ1, and the 7th row presents the performance of the state-of-the-art result-agnostic metric TfD for reference. 

From the table, the acc@10 values in the 1st to 5th rows are all higher than the last row, indicating that even trained on other projects, \techname{} still outperforms the state-of-the-art result-agnostic metric TfD. Besides, the results of \techname{} in the 1st to 5th rows are usually smaller than the ones in the 6th row, indicating that \techname{} performs worse when it is built and used on different projects. The first observation is as expected since \techname{} can learn a more accurate strategy from the FL feedback compared to the fixed heuristic-based TfD. The second observation is also as expected, since in the cross-version scenario testing data share similar features with training data, while in the cross-project scenario testing data may have different characteristics. In summary, the results demonstrate that \techname{} still performs excellently and is the state-of-the-art result-agnostic metric even in the cross-project scenario.

\subsection{RQ5. Ablation study}
In this section, we perform an ablation study to investigate the contribution of different components of \techname{}, which are the designed features, the network structure, and the training algorithm.
Besides, we systematically remove the RL component and use a weighted sum of ``split'' and ``cover'' similar to FDG~\cite{an2022fdg} to serve as a baseline to investigate the performance contribution of the entire RL component.

In particular, We design different variants of \techname{} and compare their effectiveness with \techname{} in the human-written test selection scenario to justify the design choices of our approach. The variants of \techname{} are as follows. (1) $\text{\techname{}}_{cover}$ and $\text{\techname{}}_{split}$ directly use the calculated action features $cover$ (Eq.~\ref{equ:cal_cover}) and $split$ (Eq.~\ref{equ:cal_split}) as the FDC values of tests, respectively. While \techname{} combines these two features to comprehensively predict the FDC values of tests, we investigate how these two features perform alone. (2) $\text{\techname{}}_{simpleNet}$ removes the embedding layer of the RL model, i.e., the state and the action features are directly concatenated and fed into the FDC prediction network to produce the predicted FDC value. (3) $\text{\techname{}}_{regularQ}$ uses regular Q-learning rather than double Q-learning as the training algorithm to train the RL model. (4) $\text{RLFDC}_{weight\_\alpha}$ uses a weighted sum of ``split'' and ``cover'', where $\alpha$ is the weight for ``split'' and $(1-\alpha)$ is the weight for ``cover'', i.e., $\text{RLFDC}_{weight\_\alpha}=\alpha \cdot split + (1-\alpha) \cdot cover$. We vary $\alpha$ from the discrete set $\{0.1, 0.3, 0.5, 0.7, 0.9\}$ following previous work~\cite{an2022fdg} ($\text{\techname{}}_{cover}$ and $\text{\techname{}}_{split}$ actually correspond to cases in $\text{RLFDC}_{weight\_\alpha}$ where $\alpha$ is set to 0 and 1, respectively).

Table~\ref{tab:abl} presents the performance of different variants of \techname{}. We also include the performance of \techname{} and the best result-agnostic metric TfD for reference. From the table, the designed features, the network structure, and the training algorithm all positively contribute to the effectiveness of \techname{}. Besides, the comparison between RLFDC and the simple weighted sum baselines substantiates the positive contribution of the entire RL component. In particular, $\text{\techname{}}_{cover}$ and $\text{\techname{}}_{split}$ are competitive result-agnostic FDC metrics, i.e., each of them outperforms TfD. However, both of them perform worse than \techname{}, demonstrating that both of the two features are important indicators of the FDC value, and combining them can further enhance the FDC prediction accuracy. Note that in the early stage (i.e., with a small number of selected tests) $\text{\techname{}}_{cover}$ performs better. We manually investigate some samples and find that in the early stage, since most of the selected tests are passing tests, selecting tests that cover more suspicious elements can effectively help reduce the suspicious value of non-buggy elements and thus help localize the buggy elements. However, in the late stage, covering more elements has a limited effect on more accurately localizing the fault because many ambiguity groups have been formed. Therefore, the effectiveness may be further enhanced by additionally considering how to split the ambiguous groups. The results of $\text{\techname{}}_{simpleNet}$ show the necessity of the embedding layer of the RL model. The state and action features are initially in the different hyperspaces with different scales and the embedding layer helps project them to the same space, which makes learning easier. The results of $\text{\techname{}}_{regularQ}$ show the rationality of choosing double Q-learning as our training algorithm, since regular Q-learning tends to overestimate action values and results in poor performance. 
The results of $\text{RLFDC}_{weight\_\alpha}$ show the contribution of the entire RL component. RLFDC consistently outperforms $\text{RLFDC}_{weight\_\alpha}$ with various $\alpha$. With the RL component, RLFDC automatically learns to keep a good balance between ``split'' and ``cover'' based on the current state (i.e., the number of tests and ambiguity groups). With the sense of the current state and direct feedback from fault localization results, RLFDC could learn a more precise FDC measurement and achieve better performance, as evidenced by the results in Table~\ref{tab:abl}.
In summary, different components of \techname{} all contribute to enhancing its effectiveness.

\begin{table}[tp]
\centering
\caption{acc@10 values for the test suites augmented by different variants of RLFDC}
\vspace{-0.2cm}
\begin{tabular}{l|cccc}
\toprule
\multirow{2}{*}{FDC Metric} & \multicolumn{4}{c}{\# Tests}                                       \\ \cmidrule{2-5} 
                            & \multicolumn{1}{c|}{0}  & 1            & 5            & 10           \\ \midrule
$\text{RLFDC}_{split}$            & \multicolumn{1}{c|}{65} & 92           & 133          & 163          \\
$\text{RLFDC}_{cover}$            & \multicolumn{1}{c|}{65} & \textbf{130} & 173          & 195          \\
$\text{RLFDC}_{simpleNet}$        & \multicolumn{1}{l|}{65} & 102           & 161          & 198          \\
$\text{RLFDC}_{regularQ}$        & \multicolumn{1}{l|}{65} & 91           & 163          & 183          \\ 
$\text{RLFDC}_{weight\_0.1}$                  & \multicolumn{1}{c|}{65} & 106 & 164          & 195          \\
$\text{RLFDC}_{weight\_0.3}$                  & \multicolumn{1}{c|}{65} & 97           & 156          & 202          \\
$\text{RLFDC}_{weight\_0.5}$                  & \multicolumn{1}{c|}{65} & 95           & 160          & 196          \\
$\text{RLFDC}_{weight\_0.7}$                  & \multicolumn{1}{c|}{65} & 95           & 159          & 196          \\
$\text{RLFDC}_{weight\_0.9}$                  & \multicolumn{1}{c|}{65} & 93           & 158          & 191          \\ \midrule
RLFDC                       & \multicolumn{1}{c|}{65} & 103          & \textbf{181} & \textbf{203} \\
TfD                         & \multicolumn{1}{l|}{65} & 84           & 130          & 149          \\ \bottomrule
\end{tabular}
\label{tab:abl}
\vspace{-0.4cm}
\end{table}

\section{Discussion}

\subsection{Oracle Labeling Effort}
We here discuss the oracle labeling effort, since the tests selected or generated still require manual effort to label the oracles. Note that developers are required only to verify high-level input-output relationships, which is a straightforward task for them given their intimate knowledge of the project. Initially, RLFDC can be utilized to prioritize which generated tests should be labeled first, based on their estimated impact on improving FDC. Therefore, the labeling effort could be reduced. In the worst-case scenario, each statement might need individual test coverage and differentiation from others, implying that the number of tests requiring labels would equal the number of statements. In this scenario, the overall labeling effort would scale linearly with the number of statements. However, such a scenario is uncommon, and generating such fine-grained tests would demand a highly capable test-generation tool. In such cases, developers might turn to interactive debugging techniques for additional support as discussed in Section ~\ref{sec:bug_val}. 
Manual oracle labeling could reduce the efforts of manual debugging because, with more tests, the faults could be more accurately pinpointed by SBFL which provides a preliminary focus for further investigation. 
Additionally, there are potential automated and semi-automated techniques that could assist in reducing manual efforts by automatically generating test oracles~\cite{dinella2022toga,hossain2023neural,liu2023towards}.

\subsection{Bug Validation}
\label{sec:bug_val}
SBFL only produces the probability of a program element that contains bugs. Some debugging techniques~\cite{li2018enlightened, xu2018debugging, lin2017feedback} can also help developers pinpoint the fault through interaction with humans. 
In particular, Li et al.~\cite{li2018enlightened} proposes ENLIGHTEN. ENLIGHTEN employs SBFL to form a ranking list of suspicious statements. It then seeks feedback from the developer through a series of queries about suspicious method invocations based on the SBFL results. The human feedback is encoded as extra virtual tests that can further improve SBFL results. 
Xu et al.~\cite{xu2018debugging} propose a probabilistic inference-based debugging technique. They model debugging as a probabilistic inference process where random variables represent the correctness/faultiness of each code statement and variable, and use conditional probability distributions to represent human knowledge and reasoning rules, along with program semantics. 
Lin et al.~\cite{lin2017feedback} propose a feedback-based debugging approach. The system constructs a trace model that records the program’s execution and maps out causality relationships among the steps, such as data and control dependencies. Developers can interact with the debugger by providing feedback on specific steps in the execution trace. Based on the feedback, the system iteratively refines its suspicion about which steps might be erroneous and guides the developers to likely sources of errors.

Our work can be integrated with these debugging techniques to further validate the bug location. Initially, for a given failing test, RLFDC can be employed to select/generate a certain number of additional tests to enrich debugging information. These additional tests can then serve as inputs to enhance SBFL, helping pinpoint faults with greater accuracy. The FL results can be used as guidance to the follow-up bug validation techniques, because SBFL can provide a ranked list of potentially faulty locations in the code, which serves as a preliminary focus for further investigation. For example, ENLIGHTEN~\cite{li2018enlightened} could present the enhanced SBFL results to developers and seek human feedback, potentially reducing the number of feedback iterations required. The probabilistic inference-based debugging technique~\cite{xu2018debugging} could benefit from the SBFL results by updating the fault probability of a suspicious statement according to its SBFL score, thereby more effectively recommending instances of faulty statements to users. The feedback-based debugging approach~\cite{lin2017feedback} could enhance its recommendation mechanism by utilizing SBFL results, specifically by focusing on the step in the trace with the highest SBFL score rather than relying solely on dominance relations. This strategy could potentially reduce the time required for manual checks.

\section{Threats to validity}
\textbf{Threats to internal validity} mainly come from the implementation of different metrics and the randomness caused by machine learning and evolutionary algorithms. To reduce the threats, we use the implementation of different metrics from previous work~\cite{an2022fdg} and use the published EVOSUITE artifact for EntBug~\cite{entbug} and DDU~\cite{ddu}. Besides, we use five-fold validation to validate our RL model and run the experiments for multiple times to reduce randomness. Although prior work suggests 30 repetitions as a rule of thumb~\cite{arcuri2011practical}, conducting 30 repetitions is resource-intensive and our internal analysis of the result variance from each run indicates minimal fluctuation. Furthermore, the consistently superior performance of RLFDC on tests generated by EVOSUITE and Randoop bolsters our confidence in the advantages of RLFDC.

\textbf{Threats to external validity} mainly come from the subjects, the number of selected tests, the FL technique and the test generation tool. We conduct experiments on the widely-used FL benchmark, Defects4J. We exclude some buggy programs due to deprecated faults or EVOSUITE issues, but the number of programs can still support reliable experiments. We select ten tests to augment the initial test suite following prior work~\cite{an2022fdg} to balance the efforts for human labeling and FL performance. Ochiai-agg is utilized as our FL technique since it is a representative and effective SBFL technique. We use the widely-studied EVOSUITE~\cite{fraser2011evosuite} as our test generation tool following prior work~\cite{ddu,an2022fdg,entbug} with only modification on the fitness function, which may help for generalization to other search-based test generation tools.

\textbf{Threats to construct validity} mainly come from the metrics we use. To reduce these threats, we adopt two widely-used metrics, acc@n and mAP, to measure the performance of FL, following prior work~\cite{an2022fdg}. Specifically, the former metric focuses on the highest rank of the buggy elements, while the latter metric captures the whole picture. The metrics complement each other and thus provide a comprehensive view when investigating the performance of FL.

\section{Related work}


FL is a time-consuming process in software debugging~\cite{wong2016survey}, and thus many FL techniques have been proposed~\cite{abreu2009practical,li2019deepfl,lou2021boosting} to reduce time cost for software debugging, which rely on numerous tests to diagnose faults~\cite{jin2013f3}. Therefore, researchers have put dedicated efforts in various directions around test for the purpose of FL, e.g., test augmentation~\cite{tfd,an2022fdg,entbug,ddu,artzi2010directed,jin2013f3}, test minimization~\cite{yu2008empirical,gong2012diversity,yoo2010using,alipour2016evaluating}, test purification~\cite{xuan2014test} and test prioritization~\cite{yoo2013fault,gonzalez2011prioritizing}. 

To improve the performance of FL, test augmentation techniques are proposed to generate or select more tests and enhance the diagnostic capability of the test suite. Some works focus on designing FDC metrics to evaluate the adequacy of an existing test suite for FL, and then selecting or generating more tests under the guidance of the metrics to extend the test suite. Baudry et al.~\cite{tfd} propose a test criterion, named test-for-diagnosis criterion (TfD), to evaluate the diagnosability of test cases in FL. They define dynamic basic blocks (DBB) to describe the indistinguishable statements of a program. They then implement TfD on a test optimization tool utilizing bacteriologic algorithm and aim to improve the accuracy of FL by generating more tests to reduce the size of DBB.
Campos et al.~\cite{entbug} take advantage of probability theory concepts and propose an entropy-based measurement to guide test case generation and improve FL. By extending EVOSUITE, they provide a tool named EntBug to generate new test cases.
Perez et al.~\cite{ddu} propose a new metric, named DDU, to evaluate a test suite's diagnosability for FL. Compared to previous metrics, DDU takes density, test diversity, and uniqueness into account simultaneously and is combined with EVOSUITE to generate tests to improve FL. An and Yoo~\cite{an2022fdg} propose FDG to measure the fault diagnosability gain of a test case. By leveraging the SBFL results as metric input, it locates the suspicious part of the program and effectively helps the augmentation of test suites for better FL. However, the acquisition of SBFL results needs results of the tests. Therefore, FDG can hardly be applied as a fitness function in most test generation tools. Other works exploit automated test generation from other directions rather than designing FDC metrics.
Artzi et al.~\cite{artzi2010directed} use directed concolic execution to generate tests. They aim at maximizing the path similarity between the generated tests and the failing tests to better help FL. BUGEX~\cite{robetaler2012isolating} extends EVOSUITE to generate tests to pinpoint the failure, which benefits from runtime properties like state predicates or branches. Jin et al.~\cite{jin2012bugredux} propose BugRedux to reproduce field failures of real-world programs. They then further extends BugRedux and propose a technique $F^3$~\cite{jin2013f3} to generate failing and passing executions similar to the field failure to support its customized FL technique. These techniques are test generation tools assuming there exists automated oracles, while our technique is an FDC measurement which can not only be used to guide test generation, but also be used to select tests to relive the human efforts for labelling oracles.

Besides test augmentation, some researchers study the impact of test minimization, test purification and test prioritization to FL. Yu et al.~\cite{yu2008empirical} conduct an empirical study concerning the trade-off between test suite reduction and FL effectiveness, and propose a test reduction strategy that minimizes the impact of reducing tests on FL effectiveness. Gong et al.~\cite{gong2012diversity} propose a test case selection strategy based on Diversity Maximization Speedup (DMS) to remove some test cases while maintaining the effectiveness of FL. 
Yoo et al.~\cite{yoo2010using} regard test minimization problem as a multi-objective problem. They propose a multi-objective formulation and two algorithms based on the concept of Pareto efficient to solve the test suite minimization problem. Inspired by previous work, Alipour et al.~\cite{alipour2016evaluating} propose two types of non-adequate test-case reduction metrics and evaluate the effectiveness of non-adequate reduction for test cases. In their evaluation, they find that this kind of reduction can reduce cost and maintain the similar or even better FL performance.  
Xuan et al.~\cite{xuan2014test} propose test case purification, which breaks the tests into small fractions to make full use of test oracles. The purified test cases are then used for FL and result in better FL performance. Based on the interplay between FL and test prioritization, some researchers intend to improve FL with test prioritization~\cite{yoo2013fault,gonzalez2011prioritizing} by prioritizing tests considering entropy theory or ambiguity group reduction.

\section{Conclusion}
We propose RLFDC, the first RL-based metric to measure FDC. Different from previous FDC metrics defined by hand-crafted formulas, RLFDC is a trained RL model. In the training stage, the RL model improves its measuring strategy with feedback from FL results and thus automatically learns a precise measurement. We use three scenarios to evaluate RLFDC, i.e., test selection in human-written tests, test selection in automatically generated tests, and automated test generation. The experimental results show that RLFDC has the best effectiveness in selecting high FDC tests among result-agnostic metrics and has the best effectiveness in guiding EVOSUITE to generate more high FDC tests. We also evaluate our metric in the cross-project scenario and the results show that RLFDC still outperforms the state-of-the-art result-agnostic metric, indicating its generalization capability. The results of the ablation study show that each component of RLFDC positively contributes to its effectiveness.
Our work points out a promising direction of combining RL with search-based test generation tools.

\section{Data Availability}
We make our code, data, and model checkpoints publicly available at \textbf{\url{https://github.com/yifan-CodeDir/TOSEM-RLFDC}}.

\begin{acks}
This work was supported by the National Key Research and Development Program of China under Grant No. 2023YFB4503803 and the National Natural Science Foundation of China (Grant Nos. 62372005, 62232003, 62402482).
\end{acks}

\bibliographystyle{ACM-Reference-Format}
\bibliography{ref}


\begin{thebibliography}{57}


\ifx \showCODEN    \undefined \def \showCODEN     #1{\unskip}     \fi
\ifx \showDOI      \undefined \def \showDOI       #1{#1}\fi
\ifx \showISBNx    \undefined \def \showISBNx     #1{\unskip}     \fi
\ifx \showISBNxiii \undefined \def \showISBNxiii  #1{\unskip}     \fi
\ifx \showISSN     \undefined \def \showISSN      #1{\unskip}     \fi
\ifx \showLCCN     \undefined \def \showLCCN      #1{\unskip}     \fi
\ifx \shownote     \undefined \def \shownote      #1{#1}          \fi
\ifx \showarticletitle \undefined \def \showarticletitle #1{#1}   \fi
\ifx \showURL      \undefined \def \showURL       {\relax}        \fi
\providecommand\bibfield[2]{#2}
\providecommand\bibinfo[2]{#2}
\providecommand\natexlab[1]{#1}
\providecommand\showeprint[2][]{arXiv:#2}

\bibitem[PyT(2020)]%
        {PyTorch}
 \bibinfo{year}{2020}\natexlab{}.
\newblock \bibinfo{title}{PyTorch}.
\newblock \bibinfo{howpublished}{\url{https://pytorch.org}}.
\newblock


\bibitem[Abreu et~al\mbox{.}(2009)]%
        {abreu2009practical}
\bibfield{author}{\bibinfo{person}{Rui Abreu}, \bibinfo{person}{Peter Zoeteweij}, \bibinfo{person}{Rob Golsteijn}, {and} \bibinfo{person}{Arjan~JC Van~Gemund}.} \bibinfo{year}{2009}\natexlab{}.
\newblock \showarticletitle{A practical evaluation of spectrum-based fault localization}.
\newblock \bibinfo{journal}{\emph{Journal of Systems and Software}} \bibinfo{volume}{82}, \bibinfo{number}{11} (\bibinfo{year}{2009}), \bibinfo{pages}{1780--1792}.
\newblock


\bibitem[Abreu et~al\mbox{.}(2006)]%
        {ochiai}
\bibfield{author}{\bibinfo{person}{Rui Abreu}, \bibinfo{person}{Peter Zoeteweij}, {and} \bibinfo{person}{Arjan~JC Van~Gemund}.} \bibinfo{year}{2006}\natexlab{}.
\newblock \showarticletitle{An evaluation of similarity coefficients for software fault localization}. In \bibinfo{booktitle}{\emph{2006 12th Pacific Rim International Symposium on Dependable Computing (PRDC'06)}}. IEEE, \bibinfo{pages}{39--46}.
\newblock


\bibitem[Alipour et~al\mbox{.}(2016)]%
        {alipour2016evaluating}
\bibfield{author}{\bibinfo{person}{Mohammad~Amin Alipour}, \bibinfo{person}{August Shi}, \bibinfo{person}{Rahul Gopinath}, \bibinfo{person}{Darko Marinov}, {and} \bibinfo{person}{Alex Groce}.} \bibinfo{year}{2016}\natexlab{}.
\newblock \showarticletitle{Evaluating non-adequate test-case reduction}. In \bibinfo{booktitle}{\emph{Proceedings of the 31st IEEE/ACM International Conference on Automated Software Engineering}}. \bibinfo{pages}{16--26}.
\newblock


\bibitem[An and Yoo(2022)]%
        {an2022fdg}
\bibfield{author}{\bibinfo{person}{Gabin An} {and} \bibinfo{person}{Shin Yoo}.} \bibinfo{year}{2022}\natexlab{}.
\newblock \showarticletitle{FDG: a precise measurement of fault diagnosability gain of test cases}. In \bibinfo{booktitle}{\emph{Proceedings of the 31st ACM SIGSOFT International Symposium on Software Testing and Analysis}}. \bibinfo{pages}{14--26}.
\newblock


\bibitem[Arcuri and Briand(2011)]%
        {arcuri2011practical}
\bibfield{author}{\bibinfo{person}{Andrea Arcuri} {and} \bibinfo{person}{Lionel Briand}.} \bibinfo{year}{2011}\natexlab{}.
\newblock \showarticletitle{A practical guide for using statistical tests to assess randomized algorithms in software engineering}. In \bibinfo{booktitle}{\emph{Proceedings of the 33rd international conference on software engineering}}. \bibinfo{pages}{1--10}.
\newblock


\bibitem[Artzi et~al\mbox{.}(2010)]%
        {artzi2010directed}
\bibfield{author}{\bibinfo{person}{Shay Artzi}, \bibinfo{person}{Julian Dolby}, \bibinfo{person}{Frank Tip}, {and} \bibinfo{person}{Marco Pistoia}.} \bibinfo{year}{2010}\natexlab{}.
\newblock \showarticletitle{Directed test generation for effective fault localization}. In \bibinfo{booktitle}{\emph{Proceedings of the 19th international symposium on Software testing and analysis}}. \bibinfo{pages}{49--60}.
\newblock


\bibitem[Bandyopadhyay and Ghosh(2011)]%
        {bandyopadhyay2011proximity}
\bibfield{author}{\bibinfo{person}{Aritra Bandyopadhyay} {and} \bibinfo{person}{Sudipto Ghosh}.} \bibinfo{year}{2011}\natexlab{}.
\newblock \showarticletitle{Proximity based weighting of test cases to improve spectrum based fault localization}. In \bibinfo{booktitle}{\emph{2011 26th IEEE/ACM International Conference on Automated Software Engineering (ASE 2011)}}. IEEE, \bibinfo{pages}{420--423}.
\newblock


\bibitem[Baudry et~al\mbox{.}(2006)]%
        {tfd}
\bibfield{author}{\bibinfo{person}{Benoit Baudry}, \bibinfo{person}{Franck Fleurey}, {and} \bibinfo{person}{Yves Le~Traon}.} \bibinfo{year}{2006}\natexlab{}.
\newblock \showarticletitle{Improving test suites for efficient fault localization}. In \bibinfo{booktitle}{\emph{Proceedings of the 28th international conference on Software engineering}}. \bibinfo{pages}{82--91}.
\newblock


\bibitem[Calvo and Santaf{\'e}~Rodrigo(2016)]%
        {calvo2016scmamp}
\bibfield{author}{\bibinfo{person}{Borja Calvo} {and} \bibinfo{person}{Guzm{\'a}n Santaf{\'e}~Rodrigo}.} \bibinfo{year}{2016}\natexlab{}.
\newblock \showarticletitle{scmamp: Statistical comparison of multiple algorithms in multiple problems}.
\newblock \bibinfo{journal}{\emph{The R Journal, Vol. 8/1, Aug. 2016}} (\bibinfo{year}{2016}).
\newblock


\bibitem[Campos et~al\mbox{.}(2013)]%
        {entbug}
\bibfield{author}{\bibinfo{person}{Jos{\'e} Campos}, \bibinfo{person}{Rui Abreu}, \bibinfo{person}{Gordon Fraser}, {and} \bibinfo{person}{Marcelo d'Amorim}.} \bibinfo{year}{2013}\natexlab{}.
\newblock \showarticletitle{Entropy-based test generation for improved fault localization}. In \bibinfo{booktitle}{\emph{2013 28th IEEE/ACM International Conference on Automated Software Engineering (ASE)}}. IEEE, \bibinfo{pages}{257--267}.
\newblock


\bibitem[Dinella et~al\mbox{.}(2022)]%
        {dinella2022toga}
\bibfield{author}{\bibinfo{person}{Elizabeth Dinella}, \bibinfo{person}{Gabriel Ryan}, \bibinfo{person}{Todd Mytkowicz}, {and} \bibinfo{person}{Shuvendu~K Lahiri}.} \bibinfo{year}{2022}\natexlab{}.
\newblock \showarticletitle{Toga: A neural method for test oracle generation}. In \bibinfo{booktitle}{\emph{Proceedings of the 44th International Conference on Software Engineering}}. \bibinfo{pages}{2130--2141}.
\newblock


\bibitem[Fraser and Arcuri(2011)]%
        {fraser2011evosuite}
\bibfield{author}{\bibinfo{person}{Gordon Fraser} {and} \bibinfo{person}{Andrea Arcuri}.} \bibinfo{year}{2011}\natexlab{}.
\newblock \showarticletitle{Evosuite: automatic test suite generation for object-oriented software}. In \bibinfo{booktitle}{\emph{Proceedings of the 19th ACM SIGSOFT symposium and the 13th European conference on Foundations of software engineering}}. \bibinfo{pages}{416--419}.
\newblock


\bibitem[Gong et~al\mbox{.}(2012)]%
        {gong2012diversity}
\bibfield{author}{\bibinfo{person}{Liang Gong}, \bibinfo{person}{David Lo}, \bibinfo{person}{Lingxiao Jiang}, {and} \bibinfo{person}{Hongyu Zhang}.} \bibinfo{year}{2012}\natexlab{}.
\newblock \showarticletitle{Diversity maximization speedup for fault localization}. In \bibinfo{booktitle}{\emph{Proceedings of the 27th IEEE/ACM International Conference on Automated Software Engineering}}. \bibinfo{pages}{30--39}.
\newblock


\bibitem[Gonzalez-Sanchez et~al\mbox{.}(2011)]%
        {gonzalez2011prioritizing}
\bibfield{author}{\bibinfo{person}{Alberto Gonzalez-Sanchez}, \bibinfo{person}{Rui Abreu}, \bibinfo{person}{Hans-Gerhard Gross}, {and} \bibinfo{person}{Arjan~JC van Gemund}.} \bibinfo{year}{2011}\natexlab{}.
\newblock \showarticletitle{Prioritizing tests for fault localization through ambiguity group reduction}. In \bibinfo{booktitle}{\emph{2011 26th IEEE/ACM International Conference on Automated Software Engineering (ASE 2011)}}. IEEE, \bibinfo{pages}{83--92}.
\newblock


\bibitem[Hao et~al\mbox{.}(2010)]%
        {hao2010test}
\bibfield{author}{\bibinfo{person}{Dan Hao}, \bibinfo{person}{Tao Xie}, \bibinfo{person}{Lu Zhang}, \bibinfo{person}{Xiaoyin Wang}, \bibinfo{person}{Jiasu Sun}, {and} \bibinfo{person}{Hong Mei}.} \bibinfo{year}{2010}\natexlab{}.
\newblock \showarticletitle{Test input reduction for result inspection to facilitate fault localization}.
\newblock \bibinfo{journal}{\emph{Automated software engineering}}  \bibinfo{volume}{17} (\bibinfo{year}{2010}), \bibinfo{pages}{5--31}.
\newblock


\bibitem[Hasselt(2010)]%
        {hasselt2010double}
\bibfield{author}{\bibinfo{person}{Hado Hasselt}.} \bibinfo{year}{2010}\natexlab{}.
\newblock \showarticletitle{Double Q-learning}.
\newblock \bibinfo{journal}{\emph{Advances in neural information processing systems}}  \bibinfo{volume}{23} (\bibinfo{year}{2010}).
\newblock


\bibitem[Hong et~al\mbox{.}(2015)]%
        {hong2015mutation}
\bibfield{author}{\bibinfo{person}{Shin Hong}, \bibinfo{person}{Byeongcheol Lee}, \bibinfo{person}{Taehoon Kwak}, \bibinfo{person}{Yiru Jeon}, \bibinfo{person}{Bongsuk Ko}, \bibinfo{person}{Yunho Kim}, {and} \bibinfo{person}{Moonzoo Kim}.} \bibinfo{year}{2015}\natexlab{}.
\newblock \showarticletitle{Mutation-based fault localization for real-world multilingual programs (T)}. In \bibinfo{booktitle}{\emph{2015 30th IEEE/ACM International Conference on Automated Software Engineering (ASE)}}. IEEE, \bibinfo{pages}{464--475}.
\newblock


\bibitem[Hossain et~al\mbox{.}(2023)]%
        {hossain2023neural}
\bibfield{author}{\bibinfo{person}{Soneya~Binta Hossain}, \bibinfo{person}{Antonio Filieri}, \bibinfo{person}{Matthew~B Dwyer}, \bibinfo{person}{Sebastian Elbaum}, {and} \bibinfo{person}{Willem Visser}.} \bibinfo{year}{2023}\natexlab{}.
\newblock \showarticletitle{Neural-based test oracle generation: A large-scale evaluation and lessons learned}. In \bibinfo{booktitle}{\emph{Proceedings of the 31st ACM Joint European Software Engineering Conference and Symposium on the Foundations of Software Engineering}}. \bibinfo{pages}{120--132}.
\newblock


\bibitem[Iman and Davenport(1980)]%
        {iman1980approximations}
\bibfield{author}{\bibinfo{person}{Ronald~L Iman} {and} \bibinfo{person}{James~M Davenport}.} \bibinfo{year}{1980}\natexlab{}.
\newblock \showarticletitle{Approximations of the critical region of the fbietkan statistic}.
\newblock \bibinfo{journal}{\emph{Communications in Statistics-Theory and Methods}} \bibinfo{volume}{9}, \bibinfo{number}{6} (\bibinfo{year}{1980}), \bibinfo{pages}{571--595}.
\newblock


\bibitem[Jin and Orso(2012)]%
        {jin2012bugredux}
\bibfield{author}{\bibinfo{person}{Wei Jin} {and} \bibinfo{person}{Alessandro Orso}.} \bibinfo{year}{2012}\natexlab{}.
\newblock \showarticletitle{Bugredux: Reproducing field failures for in-house debugging}. In \bibinfo{booktitle}{\emph{2012 34th International Conference on Software Engineering (ICSE)}}. IEEE, \bibinfo{pages}{474--484}.
\newblock


\bibitem[Jin and Orso(2013)]%
        {jin2013f3}
\bibfield{author}{\bibinfo{person}{Wei Jin} {and} \bibinfo{person}{Alessandro Orso}.} \bibinfo{year}{2013}\natexlab{}.
\newblock \showarticletitle{F3: Fault localization for field failures}. In \bibinfo{booktitle}{\emph{Proceedings of the 2013 International Symposium on Software Testing and Analysis}}. \bibinfo{pages}{213--223}.
\newblock


\bibitem[Jones and Harrold(2005)]%
        {jones2005empirical}
\bibfield{author}{\bibinfo{person}{James~A Jones} {and} \bibinfo{person}{Mary~Jean Harrold}.} \bibinfo{year}{2005}\natexlab{}.
\newblock \showarticletitle{Empirical evaluation of the tarantula automatic fault-localization technique}. In \bibinfo{booktitle}{\emph{Proceedings of the 20th IEEE/ACM international Conference on Automated software engineering}}. \bibinfo{pages}{273--282}.
\newblock


\bibitem[Jost(2006)]%
        {jost2006entropy}
\bibfield{author}{\bibinfo{person}{Lou Jost}.} \bibinfo{year}{2006}\natexlab{}.
\newblock \showarticletitle{Entropy and diversity}.
\newblock \bibinfo{journal}{\emph{Oikos}} \bibinfo{volume}{113}, \bibinfo{number}{2} (\bibinfo{year}{2006}), \bibinfo{pages}{363--375}.
\newblock


\bibitem[Just({[n.\,d.]})]%
        {d4jwebsite2024}
\bibfield{author}{\bibinfo{person}{René Just}.} \bibinfo{year}{[n.\,d.]}\natexlab{}.
\newblock \bibinfo{title}{Defects4J Github repository}.
\newblock \bibinfo{howpublished}{\url{https://github.com/rjust/defects4j}}.
\newblock
\newblock
\shownote{Accessed on: 2024-08-18}.


\bibitem[Just et~al\mbox{.}(2014)]%
        {just2014defects4j}
\bibfield{author}{\bibinfo{person}{Ren{\'e} Just}, \bibinfo{person}{Darioush Jalali}, {and} \bibinfo{person}{Michael~D Ernst}.} \bibinfo{year}{2014}\natexlab{}.
\newblock \showarticletitle{Defects4J: A database of existing faults to enable controlled testing studies for Java programs}. In \bibinfo{booktitle}{\emph{Proceedings of the 2014 international symposium on software testing and analysis}}. \bibinfo{pages}{437--440}.
\newblock


\bibitem[Kang et~al\mbox{.}(2024)]%
        {kang2024quantitative}
\bibfield{author}{\bibinfo{person}{Sungmin Kang}, \bibinfo{person}{Gabin An}, {and} \bibinfo{person}{Shin Yoo}.} \bibinfo{year}{2024}\natexlab{}.
\newblock \showarticletitle{A quantitative and qualitative evaluation of LLM-based explainable fault localization}.
\newblock \bibinfo{journal}{\emph{Proceedings of the ACM on Software Engineering}} \bibinfo{volume}{1}, \bibinfo{number}{FSE} (\bibinfo{year}{2024}), \bibinfo{pages}{1424--1446}.
\newblock


\bibitem[Li et~al\mbox{.}(2019)]%
        {li2019deepfl}
\bibfield{author}{\bibinfo{person}{Xia Li}, \bibinfo{person}{Wei Li}, \bibinfo{person}{Yuqun Zhang}, {and} \bibinfo{person}{Lingming Zhang}.} \bibinfo{year}{2019}\natexlab{}.
\newblock \showarticletitle{Deepfl: Integrating multiple fault diagnosis dimensions for deep fault localization}. In \bibinfo{booktitle}{\emph{Proceedings of the 28th ACM SIGSOFT international symposium on software testing and analysis}}. \bibinfo{pages}{169--180}.
\newblock


\bibitem[Li et~al\mbox{.}(2018)]%
        {li2018enlightened}
\bibfield{author}{\bibinfo{person}{Xiangyu Li}, \bibinfo{person}{Shaowei Zhu}, \bibinfo{person}{Marcelo d'Amorim}, {and} \bibinfo{person}{Alessandro Orso}.} \bibinfo{year}{2018}\natexlab{}.
\newblock \showarticletitle{Enlightened debugging}. In \bibinfo{booktitle}{\emph{Proceedings of the 40th International Conference on Software Engineering}}. \bibinfo{pages}{82--92}.
\newblock


\bibitem[Li et~al\mbox{.}(2021)]%
        {li2021fault}
\bibfield{author}{\bibinfo{person}{Yi Li}, \bibinfo{person}{Shaohua Wang}, {and} \bibinfo{person}{Tien Nguyen}.} \bibinfo{year}{2021}\natexlab{}.
\newblock \showarticletitle{Fault localization with code coverage representation learning}. In \bibinfo{booktitle}{\emph{2021 IEEE/ACM 43rd International Conference on Software Engineering (ICSE)}}. IEEE, \bibinfo{pages}{661--673}.
\newblock


\bibitem[Lin et~al\mbox{.}(2017)]%
        {lin2017feedback}
\bibfield{author}{\bibinfo{person}{Yun Lin}, \bibinfo{person}{Jun Sun}, \bibinfo{person}{Yinxing Xue}, \bibinfo{person}{Yang Liu}, {and} \bibinfo{person}{Jinsong Dong}.} \bibinfo{year}{2017}\natexlab{}.
\newblock \showarticletitle{Feedback-based debugging}. In \bibinfo{booktitle}{\emph{2017 IEEE/ACM 39th International Conference on Software Engineering (ICSE)}}. IEEE, \bibinfo{pages}{393--403}.
\newblock


\bibitem[Liu et~al\mbox{.}(2023)]%
        {liu2023towards}
\bibfield{author}{\bibinfo{person}{Zhongxin Liu}, \bibinfo{person}{Kui Liu}, \bibinfo{person}{Xin Xia}, {and} \bibinfo{person}{Xiaohu Yang}.} \bibinfo{year}{2023}\natexlab{}.
\newblock \showarticletitle{Towards more realistic evaluation for neural test oracle generation}. In \bibinfo{booktitle}{\emph{Proceedings of the 32nd ACM SIGSOFT International Symposium on Software Testing and Analysis}}. \bibinfo{pages}{589--600}.
\newblock


\bibitem[Lou et~al\mbox{.}(2020)]%
        {lou2020can}
\bibfield{author}{\bibinfo{person}{Yiling Lou}, \bibinfo{person}{Ali Ghanbari}, \bibinfo{person}{Xia Li}, \bibinfo{person}{Lingming Zhang}, \bibinfo{person}{Haotian Zhang}, \bibinfo{person}{Dan Hao}, {and} \bibinfo{person}{Lu Zhang}.} \bibinfo{year}{2020}\natexlab{}.
\newblock \showarticletitle{Can automated program repair refine fault localization? a unified debugging approach}. In \bibinfo{booktitle}{\emph{Proceedings of the 29th ACM SIGSOFT International Symposium on Software Testing and Analysis}}. \bibinfo{pages}{75--87}.
\newblock


\bibitem[Lou et~al\mbox{.}(2021)]%
        {lou2021boosting}
\bibfield{author}{\bibinfo{person}{Yiling Lou}, \bibinfo{person}{Qihao Zhu}, \bibinfo{person}{Jinhao Dong}, \bibinfo{person}{Xia Li}, \bibinfo{person}{Zeyu Sun}, \bibinfo{person}{Dan Hao}, \bibinfo{person}{Lu Zhang}, {and} \bibinfo{person}{Lingming Zhang}.} \bibinfo{year}{2021}\natexlab{}.
\newblock \showarticletitle{Boosting coverage-based fault localization via graph-based representation learning}. In \bibinfo{booktitle}{\emph{Proceedings of the 29th ACM Joint Meeting on European Software Engineering Conference and Symposium on the Foundations of Software Engineering}}. \bibinfo{pages}{664--676}.
\newblock


\bibitem[Moon et~al\mbox{.}(2014)]%
        {moon2014ask}
\bibfield{author}{\bibinfo{person}{Seokhyeon Moon}, \bibinfo{person}{Yunho Kim}, \bibinfo{person}{Moonzoo Kim}, {and} \bibinfo{person}{Shin Yoo}.} \bibinfo{year}{2014}\natexlab{}.
\newblock \showarticletitle{Ask the mutants: Mutating faulty programs for fault localization}. In \bibinfo{booktitle}{\emph{2014 IEEE Seventh International Conference on Software Testing, Verification and Validation}}. IEEE, \bibinfo{pages}{153--162}.
\newblock


\bibitem[Naish et~al\mbox{.}(2011)]%
        {naish2011model}
\bibfield{author}{\bibinfo{person}{Lee Naish}, \bibinfo{person}{Hua~Jie Lee}, {and} \bibinfo{person}{Kotagiri Ramamohanarao}.} \bibinfo{year}{2011}\natexlab{}.
\newblock \showarticletitle{A model for spectra-based software diagnosis}.
\newblock \bibinfo{journal}{\emph{ACM Transactions on software engineering and methodology (TOSEM)}} \bibinfo{volume}{20}, \bibinfo{number}{3} (\bibinfo{year}{2011}), \bibinfo{pages}{1--32}.
\newblock


\bibitem[Pacheco and Ernst(2007)]%
        {pacheco2007randoop}
\bibfield{author}{\bibinfo{person}{Carlos Pacheco} {and} \bibinfo{person}{Michael~D Ernst}.} \bibinfo{year}{2007}\natexlab{}.
\newblock \showarticletitle{Randoop: feedback-directed random testing for Java}. In \bibinfo{booktitle}{\emph{Companion to the 22nd ACM SIGPLAN conference on Object-oriented programming systems and applications companion}}. \bibinfo{pages}{815--816}.
\newblock


\bibitem[Papadakis and Le~Traon(2012)]%
        {papadakis2012using}
\bibfield{author}{\bibinfo{person}{Mike Papadakis} {and} \bibinfo{person}{Yves Le~Traon}.} \bibinfo{year}{2012}\natexlab{}.
\newblock \showarticletitle{Using mutants to locate" unknown" faults}. In \bibinfo{booktitle}{\emph{2012 IEEE Fifth International Conference on Software Testing, Verification and Validation}}. IEEE, \bibinfo{pages}{691--700}.
\newblock


\bibitem[Papadakis and Le~Traon(2015)]%
        {papadakis2015metallaxis}
\bibfield{author}{\bibinfo{person}{Mike Papadakis} {and} \bibinfo{person}{Yves Le~Traon}.} \bibinfo{year}{2015}\natexlab{}.
\newblock \showarticletitle{Metallaxis-FL: mutation-based fault localization}.
\newblock \bibinfo{journal}{\emph{Software Testing, Verification and Reliability}} \bibinfo{volume}{25}, \bibinfo{number}{5-7} (\bibinfo{year}{2015}), \bibinfo{pages}{605--628}.
\newblock


\bibitem[Perez et~al\mbox{.}(2017)]%
        {ddu}
\bibfield{author}{\bibinfo{person}{Alexandre Perez}, \bibinfo{person}{Rui Abreu}, {and} \bibinfo{person}{Arie Van~Deursen}.} \bibinfo{year}{2017}\natexlab{}.
\newblock \showarticletitle{A test-suite diagnosability metric for spectrum-based fault localization approaches}. In \bibinfo{booktitle}{\emph{2017 IEEE/ACM 39th International Conference on Software Engineering (ICSE)}}. IEEE, \bibinfo{pages}{654--664}.
\newblock


\bibitem[R{\"o}$\beta$ler et~al\mbox{.}(2012)]%
        {robetaler2012isolating}
\bibfield{author}{\bibinfo{person}{Jeremias R{\"o}$\beta$ler}, \bibinfo{person}{Gordon Fraser}, \bibinfo{person}{Andreas Zeller}, {and} \bibinfo{person}{Alessandro Orso}.} \bibinfo{year}{2012}\natexlab{}.
\newblock \showarticletitle{Isolating failure causes through test case generation}. In \bibinfo{booktitle}{\emph{Proceedings of the 2012 international symposium on software testing and analysis}}. \bibinfo{pages}{309--319}.
\newblock


\bibitem[Shamshiri et~al\mbox{.}(2015)]%
        {shamshiri2015automatically}
\bibfield{author}{\bibinfo{person}{Sina Shamshiri}, \bibinfo{person}{Ren{\'e} Just}, \bibinfo{person}{Jos{\'e}~Miguel Rojas}, \bibinfo{person}{Gordon Fraser}, \bibinfo{person}{Phil McMinn}, {and} \bibinfo{person}{Andrea Arcuri}.} \bibinfo{year}{2015}\natexlab{}.
\newblock \showarticletitle{Do automatically generated unit tests find real faults? an empirical study of effectiveness and challenges (t)}. In \bibinfo{booktitle}{\emph{2015 30th IEEE/ACM International Conference on Automated Software Engineering (ASE)}}. IEEE, \bibinfo{pages}{201--211}.
\newblock


\bibitem[Sohn and Yoo(2017)]%
        {sohn2017fluccs}
\bibfield{author}{\bibinfo{person}{Jeongju Sohn} {and} \bibinfo{person}{Shin Yoo}.} \bibinfo{year}{2017}\natexlab{}.
\newblock \showarticletitle{Fluccs: Using code and change metrics to improve fault localization}. In \bibinfo{booktitle}{\emph{Proceedings of the 26th ACM SIGSOFT International Symposium on Software Testing and Analysis}}. \bibinfo{pages}{273--283}.
\newblock


\bibitem[Van~Hasselt et~al\mbox{.}(2016)]%
        {van2016deep}
\bibfield{author}{\bibinfo{person}{Hado Van~Hasselt}, \bibinfo{person}{Arthur Guez}, {and} \bibinfo{person}{David Silver}.} \bibinfo{year}{2016}\natexlab{}.
\newblock \showarticletitle{Deep reinforcement learning with double q-learning}. In \bibinfo{booktitle}{\emph{Proceedings of the AAAI conference on artificial intelligence}}, Vol.~\bibinfo{volume}{30}.
\newblock


\bibitem[Vessey(1985)]%
        {vessey1985expertise}
\bibfield{author}{\bibinfo{person}{Iris Vessey}.} \bibinfo{year}{1985}\natexlab{}.
\newblock \showarticletitle{Expertise in debugging computer programs: A process analysis}.
\newblock \bibinfo{journal}{\emph{International Journal of Man-Machine Studies}} \bibinfo{volume}{23}, \bibinfo{number}{5} (\bibinfo{year}{1985}), \bibinfo{pages}{459--494}.
\newblock


\bibitem[Wong et~al\mbox{.}(2013)]%
        {wong2013dstar}
\bibfield{author}{\bibinfo{person}{W~Eric Wong}, \bibinfo{person}{Vidroha Debroy}, \bibinfo{person}{Ruizhi Gao}, {and} \bibinfo{person}{Yihao Li}.} \bibinfo{year}{2013}\natexlab{}.
\newblock \showarticletitle{The DStar method for effective software fault localization}.
\newblock \bibinfo{journal}{\emph{IEEE Transactions on Reliability}} \bibinfo{volume}{63}, \bibinfo{number}{1} (\bibinfo{year}{2013}), \bibinfo{pages}{290--308}.
\newblock


\bibitem[Wong et~al\mbox{.}(2016)]%
        {wong2016survey}
\bibfield{author}{\bibinfo{person}{W~Eric Wong}, \bibinfo{person}{Ruizhi Gao}, \bibinfo{person}{Yihao Li}, \bibinfo{person}{Rui Abreu}, {and} \bibinfo{person}{Franz Wotawa}.} \bibinfo{year}{2016}\natexlab{}.
\newblock \showarticletitle{A survey on software fault localization}.
\newblock \bibinfo{journal}{\emph{IEEE Transactions on Software Engineering}} \bibinfo{volume}{42}, \bibinfo{number}{8} (\bibinfo{year}{2016}), \bibinfo{pages}{707--740}.
\newblock


\bibitem[Xie et~al\mbox{.}(2022)]%
        {xie2022universal}
\bibfield{author}{\bibinfo{person}{Huan Xie}, \bibinfo{person}{Yan Lei}, \bibinfo{person}{Meng Yan}, \bibinfo{person}{Yue Yu}, \bibinfo{person}{Xin Xia}, {and} \bibinfo{person}{Xiaoguang Mao}.} \bibinfo{year}{2022}\natexlab{}.
\newblock \showarticletitle{A universal data augmentation approach for fault localization}. In \bibinfo{booktitle}{\emph{Proceedings of the 44th International Conference on Software Engineering}}. \bibinfo{pages}{48--60}.
\newblock


\bibitem[Xiong et~al\mbox{.}(2020)]%
        {xiong2020finite}
\bibfield{author}{\bibinfo{person}{Huaqing Xiong}, \bibinfo{person}{Lin Zhao}, \bibinfo{person}{Yingbin Liang}, {and} \bibinfo{person}{Wei Zhang}.} \bibinfo{year}{2020}\natexlab{}.
\newblock \showarticletitle{Finite-time analysis for double Q-learning}.
\newblock \bibinfo{journal}{\emph{Advances in neural information processing systems}}  \bibinfo{volume}{33} (\bibinfo{year}{2020}), \bibinfo{pages}{16628--16638}.
\newblock


\bibitem[Xu et~al\mbox{.}(2018)]%
        {xu2018debugging}
\bibfield{author}{\bibinfo{person}{Zhaogui Xu}, \bibinfo{person}{Shiqing Ma}, \bibinfo{person}{Xiangyu Zhang}, \bibinfo{person}{Shuofei Zhu}, {and} \bibinfo{person}{Baowen Xu}.} \bibinfo{year}{2018}\natexlab{}.
\newblock \showarticletitle{Debugging with intelligence via probabilistic inference}. In \bibinfo{booktitle}{\emph{Proceedings of the 40th International Conference on Software Engineering}}. \bibinfo{pages}{1171--1181}.
\newblock


\bibitem[Xuan and Monperrus(2014)]%
        {xuan2014test}
\bibfield{author}{\bibinfo{person}{Jifeng Xuan} {and} \bibinfo{person}{Martin Monperrus}.} \bibinfo{year}{2014}\natexlab{}.
\newblock \showarticletitle{Test case purification for improving fault localization}. In \bibinfo{booktitle}{\emph{Proceedings of the 22nd ACM SIGSOFT international symposium on foundations of software engineering}}. \bibinfo{pages}{52--63}.
\newblock


\bibitem[Yang et~al\mbox{.}(2024)]%
        {yang2024large}
\bibfield{author}{\bibinfo{person}{Aidan~ZH Yang}, \bibinfo{person}{Claire Le~Goues}, \bibinfo{person}{Ruben Martins}, {and} \bibinfo{person}{Vincent Hellendoorn}.} \bibinfo{year}{2024}\natexlab{}.
\newblock \showarticletitle{Large language models for test-free fault localization}. In \bibinfo{booktitle}{\emph{Proceedings of the 46th IEEE/ACM International Conference on Software Engineering}}. \bibinfo{pages}{1--12}.
\newblock


\bibitem[Yoo and Harman(2010)]%
        {yoo2010using}
\bibfield{author}{\bibinfo{person}{Shin Yoo} {and} \bibinfo{person}{Mark Harman}.} \bibinfo{year}{2010}\natexlab{}.
\newblock \showarticletitle{Using hybrid algorithm for pareto efficient multi-objective test suite minimisation}.
\newblock \bibinfo{journal}{\emph{Journal of Systems and Software}} \bibinfo{volume}{83}, \bibinfo{number}{4} (\bibinfo{year}{2010}), \bibinfo{pages}{689--701}.
\newblock


\bibitem[Yoo et~al\mbox{.}(2013)]%
        {yoo2013fault}
\bibfield{author}{\bibinfo{person}{Shin Yoo}, \bibinfo{person}{Mark Harman}, {and} \bibinfo{person}{David Clark}.} \bibinfo{year}{2013}\natexlab{}.
\newblock \showarticletitle{Fault localization prioritization: Comparing information-theoretic and coverage-based approaches}.
\newblock \bibinfo{journal}{\emph{ACM Transactions on software engineering and methodology (TOSEM)}} \bibinfo{volume}{22}, \bibinfo{number}{3} (\bibinfo{year}{2013}), \bibinfo{pages}{1--29}.
\newblock


\bibitem[Yu et~al\mbox{.}(2008)]%
        {yu2008empirical}
\bibfield{author}{\bibinfo{person}{Yanbing Yu}, \bibinfo{person}{James~A Jones}, {and} \bibinfo{person}{Mary~Jean Harrold}.} \bibinfo{year}{2008}\natexlab{}.
\newblock \showarticletitle{An empirical study of the effects of test-suite reduction on fault localization}. In \bibinfo{booktitle}{\emph{Proceedings of the 30th international conference on Software engineering}}. \bibinfo{pages}{201--210}.
\newblock


\bibitem[Zhang et~al\mbox{.}(2019)]%
        {zhang2019cnn}
\bibfield{author}{\bibinfo{person}{Zhuo Zhang}, \bibinfo{person}{Yan Lei}, \bibinfo{person}{Xiaoguang Mao}, {and} \bibinfo{person}{Panpan Li}.} \bibinfo{year}{2019}\natexlab{}.
\newblock \showarticletitle{CNN-FL: An effective approach for localizing faults using convolutional neural networks}. In \bibinfo{booktitle}{\emph{2019 IEEE 26th International Conference on Software Analysis, Evolution and Reengineering (SANER)}}. IEEE, \bibinfo{pages}{445--455}.
\newblock


\bibitem[Zou et~al\mbox{.}(2019)]%
        {zou2019empirical}
\bibfield{author}{\bibinfo{person}{Daming Zou}, \bibinfo{person}{Jingjing Liang}, \bibinfo{person}{Yingfei Xiong}, \bibinfo{person}{Michael~D Ernst}, {and} \bibinfo{person}{Lu Zhang}.} \bibinfo{year}{2019}\natexlab{}.
\newblock \showarticletitle{An empirical study of fault localization families and their combinations}.
\newblock \bibinfo{journal}{\emph{IEEE Transactions on Software Engineering}} \bibinfo{volume}{47}, \bibinfo{number}{2} (\bibinfo{year}{2019}), \bibinfo{pages}{332--347}.
\newblock


\end{thebibliography}

\end{document}